\documentclass{article}

\usepackage{arxiv}

\usepackage[utf8]{inputenc} 
\usepackage[T1]{fontenc}    
\usepackage{amsmath}
\usepackage{hyperref}
\usepackage{url}            
\usepackage{booktabs}       
\usepackage{amsfonts}       
\usepackage{nicefrac}       
\usepackage{microtype}      
\usepackage{cleveref}       
\usepackage{graphicx}
\usepackage{subcaption}
\usepackage{gensymb}
\usepackage{lipsum}
\usepackage[numbers,super,sort&compress]{natbib}
\usepackage{doi}
\usepackage{algorithm}
\usepackage{algpseudocode}
\usepackage{booktabs}
\usepackage[table]{xcolor}

\definecolor{tabfirst}{rgb}{1, 0.7, 0.7} 
\definecolor{tabsecond}{rgb}{1, 0.85, 0.7} 
\definecolor{tabthird}{rgb}{1, 1, 0.7} 

\hypersetup{
    colorlinks=true,     
    urlcolor=blue,
    citecolor=black,
    linkcolor=black,
    }

\algtext*{EndIf} 
\algtext*{EndWhile} 
\algtext*{EndFor} 
\algtext*{EndFunction} 

\newcommand{\name}{{\sc MHNpath}}

\title{A User-Tunable Machine Learning Framework for Step-Wise Synthesis Planning}

\date{}

\author{%
 Shivesh Prakash$^{1}$\href{https://orcid.org/0009-0007-4120-0921}{\includegraphics[width=10pt]{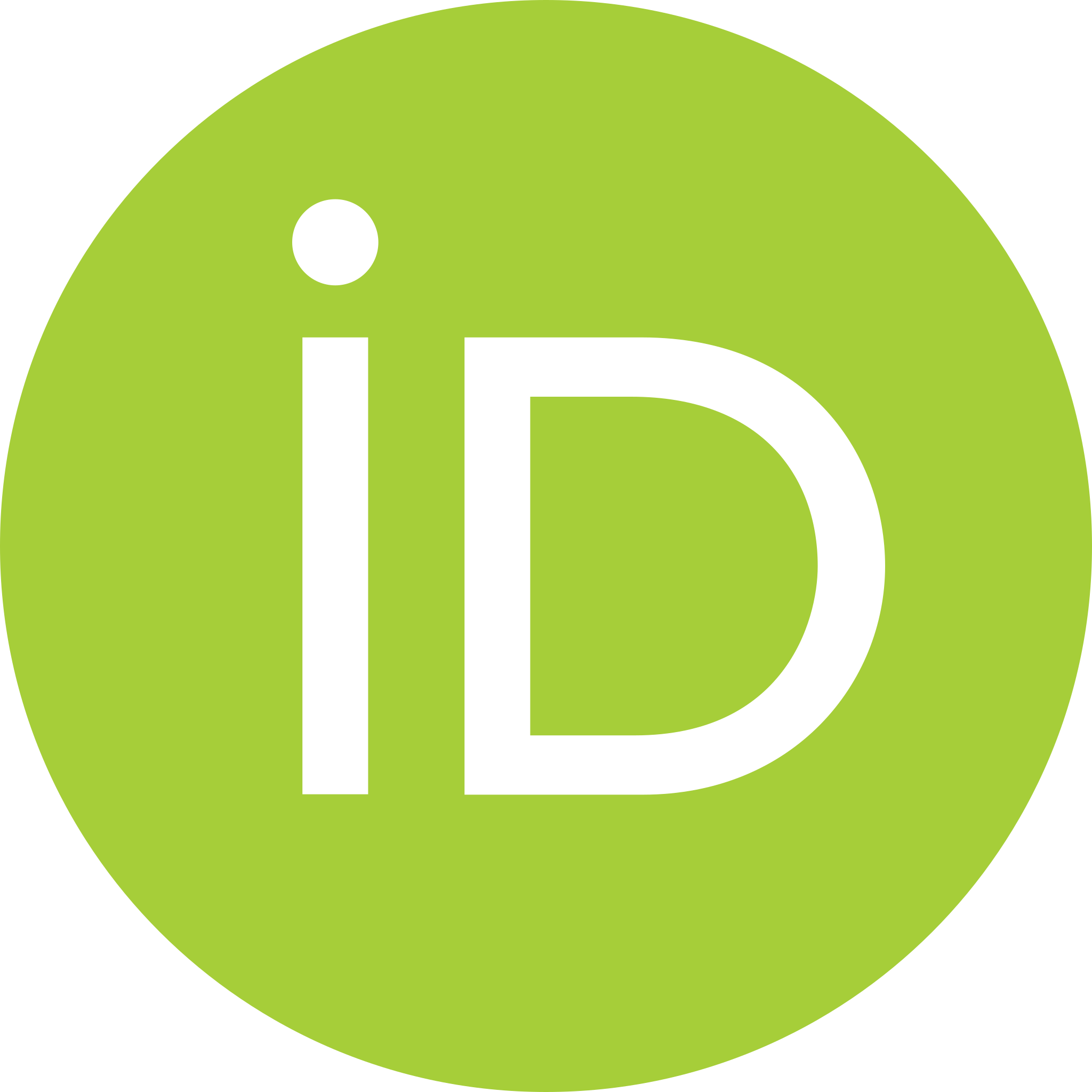}}\quad 
 Nandan Patel$^{4}$\href{https://orcid.org/0009-0003-5660-5658}{\includegraphics[width=10pt]{figures/ORCID.png}}\quad 
 Hans-Arno Jacobsen$^{1, 2}$\href{https://orcid.org/0000-0003-0813-0101}{\includegraphics[width=10pt]{figures/ORCID.png}}\quad 
 Viki Kumar Prasad$^{2, 3, 4 *}$\href{https://orcid.org/0000-0003-0982-3129}{\includegraphics[width=10pt]{figures/ORCID.png}} \\
 {\normalfont $^1$Department of Computer Science, University of Toronto, 40 St George St, Toronto, ON M5S 2E4} \\
 {\normalfont $^2$The Edward S. Rogers Sr. Department of Electrical \& Computer Engineering, University of Toronto,} \\ {\normalfont 10 King's College Rd, Toronto, ON M5S 3G8} \\
 {\normalfont $^3$Data Sciences Institute, University of Toronto, 700 University Ave 10th floor, Toronto, ON M7A 2S4} \\
 {\normalfont $^4$Current affiliation: Department of Chemistry, University of Calgary, 2500 University Drive NW, Calgary, AB T2N 1N4} \\
 $^{*}$\texttt{vikikumar.prasad@ucalgary.ca}
}

\renewcommand{\headeright}{Prakash et al.}
\renewcommand{\undertitle}{}
\renewcommand{\shorttitle}{\name}

\hypersetup{
pdftitle={\name: A Deep Learning Framework for Step-Wise Synthesis Planning via Data-Driven Exploration},
pdfsubject={cs.AI, cs.LG, q-bio.QM},
pdfauthor={Shivesh~Prakash},
pdfkeywords={First keyword, Second keyword, More},
}

\begin{document}
\maketitle

\begin{abstract}

We introduce \name, a machine learning-driven retrosynthetic tool designed for computer-aided synthesis planning. Leveraging modern Hopfield networks and novel comparative metrics, \name\ efficiently prioritizes reaction templates, improving the scalability and accuracy of retrosynthetic predictions. The tool incorporates a tunable scoring system that allows users to prioritize pathways based on cost, reaction temperature, and toxicity, thereby facilitating the design of greener and cost-effective reaction routes. We demonstrate its effectiveness through case studies involving complex molecules from ChemByDesign, showcasing its ability to predict novel synthetic and enzymatic pathways. Furthermore, we benchmark \name\ against existing frameworks using the PaRoutes dataset, achieving a solution rate of 85.4\% and replicating 69.2\% of experimentally validated ``gold-standard'' pathways. Our case studies show that the tool can generate shorter, cheaper, moderate-temperature routes employing green solvents, as exemplified by molecules such as dronabinol, arformoterol, and lupinine. 

\end{abstract}

\vspace{\baselineskip}


\section*{1 \hspace{0.25cm} Introduction}

The integration of machine learning (ML) into chemical synthesis has transformed the field of computer-aided synthesis planning (CASP), providing chemists with powerful tools to design and execute synthetic routes more efficiently. Traditional retrosynthetic analysis, which relies heavily on expert intuition and experience \cite{Szymkuc2016, PENSAK1977}, is increasingly being augmented by data-driven approaches \cite{D4DD00120F} to break down complex molecules into simpler precursors using ML techniques to predict viable synthetic pathways \cite{szymkuc2016computer, levin2022merging}. These advancements are particularly crucial in addressing the growing complexity of modern synthetic challenges, where exploring vast chemical spaces catalyzes the discovery of novel molecules. Examples of such challenges include the synthesis of complex natural products with multiple stereocenters \cite{nicolaou2008molecules} and the need for highly selective functional group transformations in pharmaceutical development \cite{brown2016analysis}.

In CASP, clean chemical data is the foundation for computationally designing synthetic routes. This data enables training of a predictive model that finds precursors while an efficient search algorithm navigates the expansive chemical space to propose feasible retrosynthetic pathways. State-of-the-art ML models for CASP can be categorized into template-based and template-free approaches where templates are generalized representations that encapsulate the core chemical transformation patterns inherent to each reaction. Complementing the ML model and search algorithm is an integrated scoring system that evaluates routes based on multiple criteria, such as cost and reaction temperature. Together, these three components establish a cohesive framework that underpins modern computer-aided synthesis planning.

Recent advancements in ML, including the development of deep neural networks and transformer-based models, have significantly enhanced the predictive capabilities of template-free methods in CASP \cite{hart1968formal, schwaller2019molecular, D1SC02362D, tetko2020state, Irwin_2022}. Transformer models have emerged as a powerful tool in retrosynthetic prediction due to their ability to handle sequence-to-sequence tasks effectively \cite{D5DD00153F}. The Molecular Transformer model \cite{schwaller2019molecular}, and its variants \cite{D1SC02362D, tetko2020state, Irwin_2022}, for instance, has demonstrated remarkable accuracy in predicting both reactants and reaction conditions by treating retrosynthesis as a language translation problem from products to reactants. These models leverage self-attention mechanisms to capture complex dependencies within chemical reactions, making them highly effective for single-step retrosynthesis \cite{WANG2021129845,andronov2025specbeam}. Some graph-based methods \cite{bradshaw2018a, sacha2021molecule} predict the products from reactants in an auto-regressive fashion, while others \cite{C8SC04228D, qian2020integrating, pmlr-v139-bi21a} predict the products by anticipating the final states of bonds or electrons. However, the reliance of template-free approaches on extensive labeled reaction data, computational resources, and single-step focus has been a significant limitation in CASP.

On the other hand, template-based approaches have also been fundamental in CASP.\cite{corey1985computer} These approaches directly apply predefined reaction rules to fragment target molecules into simpler structures, making them more interpretable and chemically intuitive compared to template-free approaches. Segler et al.\cite{segler2017neural} and Coley et al.\cite{coley2017prediction} were among the first ones to use automatically extracted rules to predict outcomes of organic reactions using neural networks while Zhang et al. \cite{zhang2025groupretro} extend it by also continually evolving them and Rho et al. \cite{roh2025higherlevel} give a global outlook by effectively abstracting the detailed substructures. Chen et al. \cite{chen2022generalized} proposed a chemistry-motivated graph neural network that uses templates to describe the net changes in electron configuration between the reactants and products. Recent advancements, such as the use of Modern Hopfield Networks for template prediction, have improved the performance of template-based CASP tools \cite{seidl2022improving}. Modern Hopfield Networks allow for the efficient prioritization of reaction templates by leveraging associative memory mechanisms, significantly improving the speed and reliability of retrosynthetic planning. Nevertheless, template-based methods often struggle with the rigidity of predefined rules, analyzed by Choe et al. \cite{choe2025crosstalk}, which can limit their ability to propose novel synthetic routes or adapt to emerging reaction types.

Computational retrosynthesis has traditionally been treated as a tree search problem, where each step involves searching for chemically feasible precursors to derive the product molecule. Initial works have used greedy search \cite{levin2022merging}, but more recent efforts have adopted Monte Carlo tree search \cite{D0SC04184J} and A$^*$-like algorithms \cite{chen2020retro, yu2024doubleendedsynthesisplanninggoalconstrained}. These search methodologies attempt to optimize the exploration of synthetic routes by balancing the exploitation of high-confidence reaction pathways with the exploration of novel or underutilized transformations. However, their effectiveness is contingent on the quality and comprehensiveness of the reaction databases they utilize.

In addition to the different components of CASP, the integration of biocatalysis in computational retrosynthetic planning has gained attention for its potential to enhance sustainability. Biocatalysis leverages enzymatic transformations to achieve highly selective reactions under mild conditions, reducing the environmental footprint of chemical synthesis \cite{finnigan2021retrobiocat}. Incorporating enzymatic steps into synthetic pathways is particularly advantageous for pharmaceutical and fine chemical industries, where stringent purity and selectivity requirements are critical \cite{B713736M, Horvath2007}. Tools like RetroBioCat \cite{finnigan2021retrobiocat} have demonstrated the ability to design biocatalytic cascades that complement synthetic routes by introducing enzymatic steps that are often more selective and environmentally friendly. However, even these specialized tools are limited by the availability of comprehensive enzymatic reaction data and their focus on narrow reaction types.

\begin{figure}[h!]

\begin{subfigure}{\textwidth}
    \centering
    \includegraphics[width=0.91\textwidth]{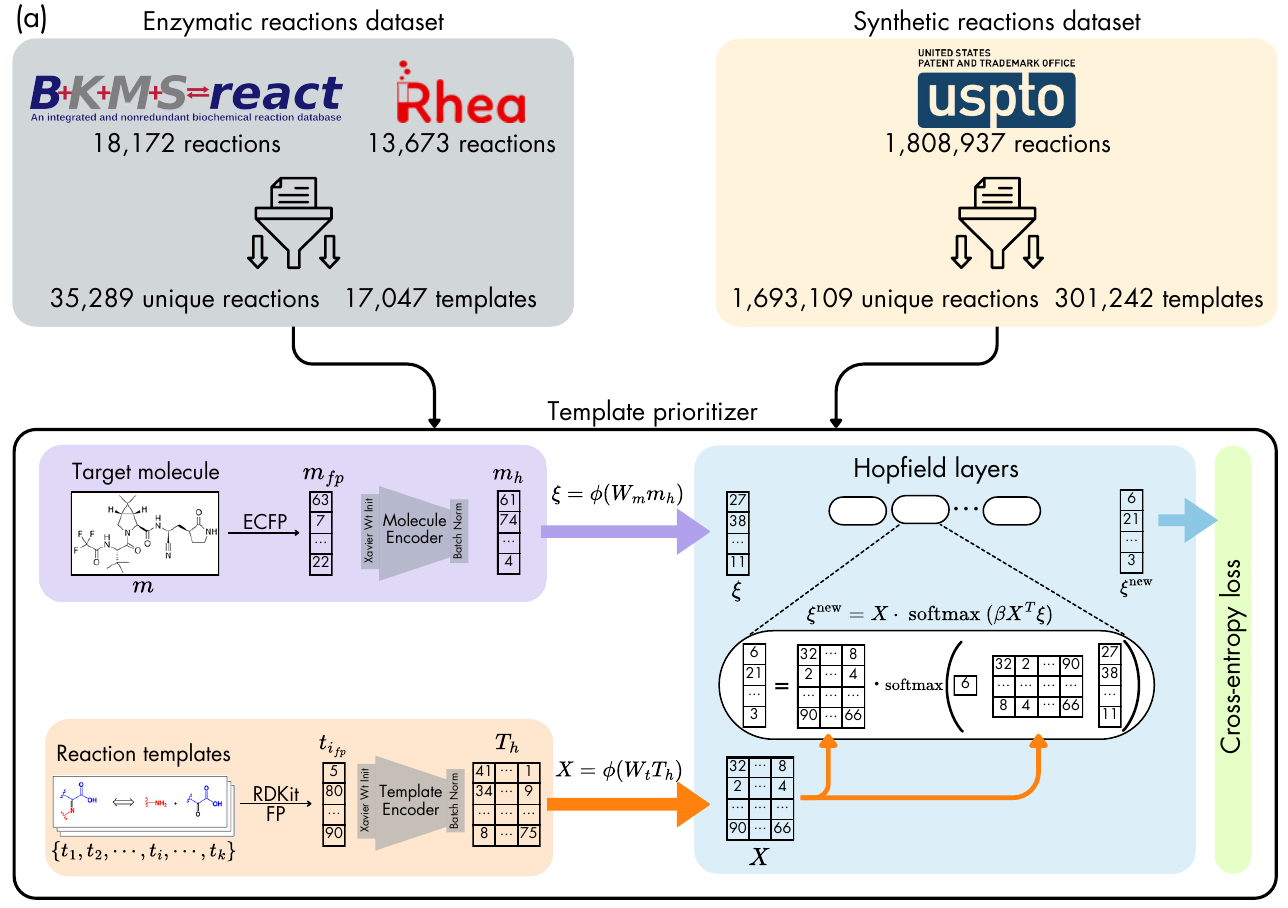}
    \label{fig:1a}
\end{subfigure}

\begin{subfigure}{\textwidth}
    \centering
    \includegraphics[width=0.91\textwidth]{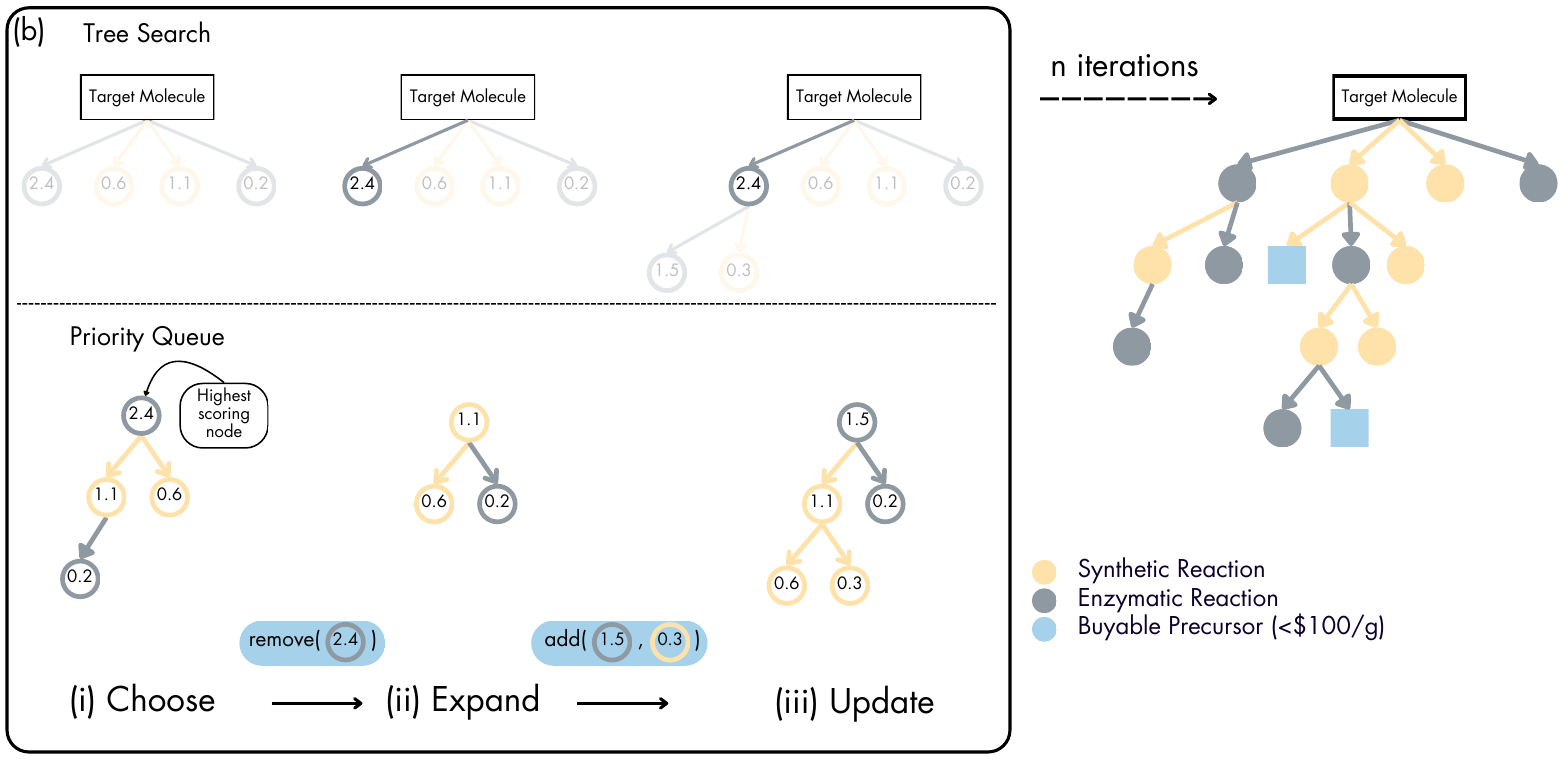}
    \label{fig:1b}
\end{subfigure}

\caption{\textbf{(a) Data processing and model architecture.} Enzymatic and synthetic databases were cleaned, and reaction templates were algorithmically extracted from them. A Modern Hopfield Network-based template prioritizer was trained to predict the reaction template associated with each product molecule in the reaction database.\\ \textbf{(b) Tree search methodology.} Multi-modal-guided-global-greedy tree search approach was used to explore the retrosynthetic search space. The precursors are denoted as grey and yellow circles for the synthetic and enzymatic reactions, respectively. The nodes are assigned a score based on a methodology that promotes low cost, the use of green solvents, and moderate-temperature reactions. A priority queue is maintained to track the highest-scoring node at all times. The highest-scoring node from the queue is explored. The queue and tree are updated iteratively until the tree is fully explored to a certain depth, or we reach a time limit.}
\label{fig:1}        
\end{figure}

In this work, we introduce \name, an ML-driven retrosynthetic tool designed to help chemists explore greener, cost-effective, and moderate-temperature synthetic pathways. First, we develop a robust template-prioritization model based on Modern Hopfield Networks \cite{ramsauer2020hopfield}, enhanced with Xavier initialization, dropout, and L2 regularization to improve stability and generalization. We additionally introduce two new evaluation metrics that enable a more comprehensive comparison of template prioritizers. Second, we implement a user-tunable scoring system that integrates practical constraints and include precursor cost, reaction temperature, and solvent toxicity, allowing chemists to steer the search toward more sustainable and operationally feasible routes. Third, we demonstrate \name’s effectiveness through case studies from PaRoutes \cite{D2DD00015F} and ChemByDesign \cite{draghici2012chemistry}, showcasing its ability to identify viable, interpretable multi-step pathways. Together with its curated enzymatic and synthetic template datasets and a global greedy tree-search algorithm, \name\ provides a flexible and data-efficient framework that addresses the limitations of existing CASP tools, particularly their reliance on extensive training and rigid template-based systems. The following sections will detail the methodology behind \name, including the data processing pipeline (\textit{Section 2.1}), the implementation of the Modern Hopfield Network-based template prioritizer (\textit{Section 2.2}), and the tree search methodology and custom scoring employed for retrosynthetic exploration (\textit{Sections 2.3 \& 2.4}), as outlined in \Cref{fig:1}. Furthermore, we will present results (\textit{Section 3}) comparing \name\ against existing CASP tools and provide case studies demonstrating its practical applications.

\newpage

\section*{2 \hspace{0.25cm} Methods}

The \name\ framework is presented in \Cref{fig:1}. A modified Modern Hopfield Network \cite{ramsauer2020hopfield} based architecture by Seidl et al. \cite{seidl2022improving} is adapted. Each prioritizer (also referred to as model) accepts a target molecule as input and predicts a ranked list of the most applicable reaction templates, which are applied to the target molecule to get precursors (nodes). The individual scores for precursor cost, reaction temperature, and solvent toxicity are calculated and summed up using user-tunable weights. The precursors are further explored in a recursive manner using a global greedy tree search algorithm where the best-scoring unexplored node from the entire tree is explored first. \Cref{fig:1}(b) gives an example of the multi-modal-guided global greedy tree search approach.

\subsection*{2.1 \hspace{0.2cm} Dataset processing}

Our study leveraged two comprehensive training datasets to develop the predictive component of the \name\ framework. Additionally, specific external datasets were curated for benchmarking and evaluation tasks. In this context, we selected training data that provided a comprehensive representation of both enzymatic and synthetic reactions. The training dataset of enzymatic reactions was compiled from the BKMS\cite{lang2011bkm} dataset and the Rhea\cite{bansal2022rhea} database, initially containing about 68,000 reactions. Following a rigorous preprocessing procedure that involved removing duplicates, invalid SMILES representations, and unbalanced reactions, we obtained 35,289 unique and valid reaction SMILES. From this refined enzymatic subset, we extracted 17,047 reaction templates using the RDKit\cite{Landrum2016RDKit2016_09_4} library. This subset was used to train a single \textit{enzymatic template prioritizer}, utilizing an 80:10:10 random split for training, validation, and testing, respectively.

We also used a dataset of  synthetic organic reactions for training which was sourced from the USPTO\cite{Lowe2017} dataset. Starting with 1,808,937 reactions and applying a series of preprocessing steps resulted in 1,693,109 refined reactions, from which 301,242 reaction templates were extracted in the synthetic subset. To manage computational resources and memory overheads due to the large number of data points in the synthetic subset, we employed a scaffold split technique using Datamol \cite{hadrien_mary_2024_10535844} to ensure high chemical diversity while dividing the subset further into five equally sized subsets. This partitioning strategy allowed us to train five distinct \textit{synthetic template prioritizers} (one per divided synthetic subset). Each divided subset underwent an 80:10:10 random split for model development. During inference, the predictions from these five models are aggregated using an ensemble approach. This strategy significantly enhances robustness with a single \textit{synthetic template prioritizer} achieving a top-1 accuracy of 36.6\% on its own test set and the ensemble method reaching a top-1 accuracy of 42.2\% on the combined test set. In this context, top-1 accuracy denotes the model's ability to accurately prioritize the correct ground-truth template as the number one rank in the prediction list.

Once the predictive components of {\name} were developed, we rigorously assessed the framework's performance in Sections 3.2 and 3.3 using three tasks as described below. These tasks required curating distinct datasets from the literature and provide unbiased performance metrics for systems not present in the training set.

\begin{itemize}
    \item \textbf{PaRoutes benchmark:} To evaluate pathway reconstruction against patent literature, we utilized the PaRoutes\cite{D2DD00015F} dataset. We applied a scaffold split to identify the ten most commonly occurring scaffolds within PaRoutes. From the clustering analysis, we selected 130 diverse target molecules to test our framework's ability to replicate and improve upon patent-derived "gold-standard" routes.
    \item \textbf{Novelty assessment (ChemByDesign):} To test generalization to unseen data, we selected 5 target molecules from ChemByDesign \cite{draghici2012chemistry}. We specifically filtered for pathways published after 2021 with fewer than six steps. This filtering ensured that the ground-truth reactions were not present in our USPTO training data (which was last updated in September 2016), providing a test of zero-shot pathway prediction.
    \item \textbf{Hybrid comparative set:} To benchmark against existing hybrid planners, we selected specific complex targets (e.g., dronabinol, arformoterol, 4-ethenyl-2-fluorophenol) used in prior studies by Levin et al. \cite{levin2022merging} and RetroBioCat \cite{finnigan2021retrobiocat}. These molecules were manually curated to allow for a direct qualitative comparison of pathway length, cost, and enzyme usage.
\end{itemize}

\subsection*{2.2 \hspace{0.2cm} Model development}

We utilize the Modern Hopfield Network \cite{ramsauer2020hopfield} based template prioritization architecture for our study. This architecture, initially introduced by Seidl et al. \cite{seidl2022improving}, consists of three main components: a molecule encoder, a reaction template encoder, and one or more stacked or parallel Hopfield layers. 

The molecule encoder function, denoted as $h_{w}^{m}(\cdot)$, learns a relevant representation for the input molecule $m$. We utilize a fingerprint-based approach, specifically the extended connectivity fingerprint (ECFP), coupled with a fully connected neural network with weights $w$. This encoder maps a molecule to a dense representation $\mathbf{m}_h = h_{w}^m (m)$ of dimension $d_m$. 

Similarly, the reaction template encoder function, denoted as $h_{v}^t (\cdot)$, learns relevant representations of reaction templates. We employ a fully connected neural network with RDKit template fingerprints as input. The function is applied to the set of all templates $\mathcal{T}$, and the resulting vectors are concatenated column-wise into a matrix $\mathbf{T}_h = h_{v}^t (\mathcal{T})$ with shape $(d_t, K)$, where $K$ is the number of templates. Both encoders utilize Xavier Initialization \cite{pmlr-v9-glorot10a} and Batch Normalization \cite{ioffe2015batch} to improve convergence and prevent vanishing gradients.

The core of the model consists of Hopfield layers, denoted as $g(\cdot, \cdot)$, which associate the molecule with the memory of templates. To perform the retrieval, the encoded molecule $\mathbf{m}_h$ and the template matrix $\mathbf{T}_h$ are projected into a common associative space to form the state pattern $\boldsymbol{\xi}$ and the stored patterns $\mathbf{X}$, respectively. The Hopfield layer then updates the molecule representation via an attention-like mechanism. The update rule yielding the new state $\boldsymbol{\xi}_{\text{new}}$ is defined as:

\begin{equation}
    \boldsymbol{\xi}_{\text{new}} = \mathbf{X} \mathbf{p} = \mathbf{X} \cdot \mathrm{softmax}\left(\beta \mathbf{X}^\top \boldsymbol{\xi}\right)
\end{equation}

\noindent where $\beta$ is a learnable scaling parameter (inverse temperature), $\mathbf{p}$ is the vector of associations (probabilities) over the templates, and $\mathrm{softmax}$ is applied column-wise.

For the loss function, given a training pair $(m, t)$ and the set of all templates $\mathcal{T}$, the model aims to maximize the probability assigned to the correct template $t$. We employ the negative log-likelihood as the loss function, optimizing parameters using stochastic gradient descent via the AdamW optimizer. To prevent overfitting, we apply dropout and L2 regularization. Additionally, a post-processing fingerprint-based substructure screen is used during inference to filter out chemically non-applicable templates.

We train two sets of models on our datasets: the \textit{synthetic template prioritizer} (consisting of five models) and the \textit{enzymatic template prioritizer}. We utilized PyTorch \cite{pytorch2019} version 1.9.0 and Python version 3.8 to implement the model. These models take a target molecule as input and rank the most applicable rules within the collected dataset. Given the same inputs, the outputs from the five \textit{synthetic template prioritizer} are taken individually, and a final ranking of the templates is collated using the highest predicted score. Furthermore, to fine-tune the models for optimal performance, hyperparameter tuning was performed using a one-factor-at-a-time (OFAT) approach. We tuned number of epochs, concatenation threshold, dropout rate, learning rate and some hopfield parameters, more information on the parameters and the values chosen can be found in Supplementary Method 1, 2 and Supplementary Table 1, 2. Each experiment involved modifying a single hyperparameter from a baseline configuration while keeping all other settings constant. The performance of each model was assessed using an evaluation score, with lower validation loss indicating better performance. 

\subsection*{2.3 \hspace{0.2cm} Searching algorithm}

We employ an A$^*$-inspired \cite{Hart1968} global greedy tree search approach to explore synthesis pathways for a target molecule. This method leverages the template prioritizer models to identify the most applicable reaction rules, which are then applied to the target molecule to derive precursor molecules. In this search framework, precursor molecules are represented as nodes in a tree, while reactions and their associated conditions serve as edges. However, not all precursors may be readily available or affordable. We define "buyable'' as being available for purchase at a cost of under \$100/g. The search algorithm proceeds as follows:
\begin{itemize}
    \item \textbf{Initialization:} We begin by initializing the search with the target molecule as the \textit{start\_node}. This node is added to a priority queue with high priority.
    \item \textbf{Global greedy search:} The main search loop iterates until the priority queue is empty. At each iteration, the node with the highest priority is popped from the queue.
    \item \textbf{Goal check:} If the node meets the goal criteria (e.g., low cost or maximum depth), the algorithm continues to the next iteration.
    \item \textbf{Rule application:} For the current node, we find applicable transformation rules and apply each rule to generate new chemical structures. The properties of these new structures (such as cost, temperature, and solvent score) are calculated.
    \item \textbf{Node insertion:} Each new structure is encapsulated in a \textit{new\_node}, which is added to the current node’s subtrees and inserted into the priority queue based on its score.
    \item \textbf{Scoring:} Nodes are scored based on a user-tunable criterion that includes precursor cost, reaction temperature, and solvent toxicity. The highest-scoring nodes are explored first.
    \item \textbf{Termination:} The search continues iteratively until buyable precursors are found, the tree is fully explored up to a specified maximum depth, or the allotted search time is exhausted.
\end{itemize}

\Cref{alg:global_greedy_tree_search} outlines the global greedy tree search algorithm.

\begin{algorithm}
\caption{Global Greedy Tree Search for Chemical Pathways}
\label{alg:global_greedy_tree_search}
\begin{algorithmic}[1]

\State \textbf{Initialize}
\State $N_0 \gets$ initial chemical structure with cost $C_0$
\State $PQ \gets \{\}$ \Comment{Priority queue}
\State \textbf{add} $N_0$ to $PQ$ with score $P_0=\infty$

\While{$PQ \neq \emptyset$}
    \State $N \gets$ \textbf{pop} node with highest score from $PQ$
    
    \If{goal\_check($N$)}
        \State \textbf{continue}
    \EndIf
    
    \State $R \gets$ find\_rules($N$)
    
    \For{$r \in R$}
        \State $N' \gets$ apply\_rule($r$, $N$)
        \State $C' \gets$ calculate\_cost($N'$)
        \State $T' \gets$ calculate\_temp($N'$)
        \State $S' \gets$ calculate\_solvent\_score($N'$)
        \State $P' \gets$ calculate\_score($C', T', S'$)
        
        \State \textbf{create} new node $N'$ with $C', T', S'$
        \State \textbf{add} $N'$ to subtrees of $N$
        \State \textbf{insert} $N'$ into $PQ$ with priority $P'$
    \EndFor
\EndWhile

\Function{find\_rules}{$N$}
    \State \textbf{return} predicted rules for $N$
\EndFunction

\Function{apply\_rule}{$r, N$}
    \State \textbf{return} new chemical structure from applying $r$ to $N$
\EndFunction

\Function{calculate\_priority}{$C', T', S'$}
    \State $P' \gets f(C', T', S')$ \Comment{Compute score}
    \State \textbf{return} $P'$
\EndFunction

\Function{goal\_check}{$N$}
    \State \textbf{return} \textbf{True} if $N$ meets goal conditions (low $C$ or max depth), else \textbf{False}
\EndFunction

\State \textbf{Output}
\State Print resulting tree of chemical reactions

\end{algorithmic}
\end{algorithm}

\subsection*{2.4 \hspace{0.2cm} Scoring system}

We introduce a user-tunable scoring methodology to evaluate synthesis pathways. In this approach individual scores for three features are obtained and summed based on user-tunable weights, allowing users to prioritize features according to their preferences during the pathway search. The first feature of the score is the cost. We utilize the ASKCOS\cite{coley2019robotic} buyable dataset, Molport \cite{MolPort2020}, Mcule \cite{Kiss2012}, and Chemspace \cite{chemspace} APIs with the ChemPrice\cite{Sorkun2024} library to determine the cost and availability of molecules. A molecule is considered buyable if it can be purchased for less than \$100/g \cite{levin2022merging, sterling2015zinc}. Users have the flexibility to adjust this threshold based on their objectives. The score is calculated as $- \frac{\text{price}}{500}$, as we aim to maximize the score, thus exploring reactions involving cheaper precursors first. The value 500 is chosen as a normalizer because we define anything costing above \$500/g as non-buyable; the user can modify this threshold. These threshold aligns with the affordability criteria for research-grade building blocks utilized in major chemical databases \cite{sterling2015zinc}. 

The second feature of the score is the reaction temperature. We employ a CASP tool developed by Gao et al. \cite{gao2018using} to predict the temperature at which a reaction might occur. This prediction is essential for exploring pathways where reactions may not have been previously documented or experimented on. To find the predicted temperature, we take the weighted average of the top ten predictions made by the tool, a strategy suggested by the authors to improve accuracy. The score is calculated as $- \frac{\text{temperature}}{300}$, as we aim to maximize the score, thus exploring reactions involving lower temperatures first. The value 300 is chosen as a normalizer because we define any reaction requiring over 300$^\circ$ C as non-practical; this cutoff reflects the upper operating limits of standard synthetic laboratory equipment \cite{razzaq2010continuous}.

The third feature of the score is the solvent and reagent greenness-toxicity score. We use the same CASP tool developed by Gao et al. \cite{gao2018using} to predict the solvent and reagent required for a reaction. We curate a toxicity and greenness dataset, assigning a score of -1 for toxic molecules, 0 for neutral molecules, and +1 for green or natural molecules. We use the ACS Solvent Selection Guide \cite{ACSGreenChemistryInstitute2011} to classify 100 commonly used solvents. We also use the SuperNatural 3.0 dataset \cite{10.1093/nar/gkac1008}, containing 350,000 natural products, and the T3DB dataset \cite{10.1093/nar/gku1004} to classify 4,000 toxic molecules.

\section*{3 \hspace{0.25cm} Results and Discussion}

In this section, we evaluate the performance of the \name\ framework in prioritizing reaction templates and generating feasible retrosynthetic pathways. We benchmark our predictive models against established baselines and analyze the practical utility of the proposed scoring system in guiding pathway discovery.

\subsection*{3.1 \hspace{0.2cm} Model performance}

\definecolor{tabfirst}{rgb}{1, 0.7, 0.7} 
\definecolor{tabsecond}{rgb}{1, 0.85, 0.7} 
\definecolor{tabthird}{rgb}{1, 1, 0.7} 

\begin{table*}[h!]
\centering
\setlength{\tabcolsep}{4pt} 
\begin{tabular}{lcccc|cccc|cccc}
\textbf{} & \multicolumn{4}{c}{\textbf{Lit. Rule Accuracy}} & \multicolumn{4}{c}{\textbf{Avg. Applicable Rules}} & \multicolumn{4}{c}{\textbf{Presence of Applicable Rule}} \\
\cmidrule(lr){2-5} \cmidrule(lr){6-9} \cmidrule(lr){10-13}
\textbf{Model} & \textbf{T1} & \textbf{T10} & \textbf{T50} & \textbf{T100} & \textbf{T1} & \textbf{T10} & \textbf{T50} & \textbf{T100} & \textbf{T1} & \textbf{T10} & \textbf{T50} & \textbf{T100} \\
\midrule
DNN \cite{levin2022merging} 
 & \cellcolor{tabthird}0.100 & \cellcolor{tabthird}0.312 & \cellcolor{tabthird}0.442 & \cellcolor{tabthird}0.485 
 & \cellcolor{tabthird}0.128 & \cellcolor{tabthird}0.898 & \cellcolor{tabthird}4.876 & \cellcolor{tabthird}14.005 
 & \cellcolor{tabthird}0.128 & \cellcolor{tabsecond}0.539 & \cellcolor{tabfirst}0.900 & \cellcolor{tabsecond}0.973 \\
MHNreact \cite{seidl2022improving} 
 & \cellcolor{tabsecond}0.181 & \cellcolor{tabsecond}0.566 & \cellcolor{tabsecond}0.760 & \cellcolor{tabsecond}0.809 
 & \cellcolor{tabsecond}0.199 & \cellcolor{tabsecond}1.004 & \cellcolor{tabsecond}5.647 & \cellcolor{tabsecond}16.388 
 & \cellcolor{tabsecond}0.199 & \cellcolor{tabthird}0.531 & \cellcolor{tabthird}0.861 & \cellcolor{tabthird}0.967 \\
Ours (Enz) 
 & \cellcolor{tabfirst}0.183 & \cellcolor{tabfirst}0.576 & \cellcolor{tabfirst}0.763 & \cellcolor{tabfirst}0.814 
 & \cellcolor{tabfirst}0.201 & \cellcolor{tabfirst}1.076 & \cellcolor{tabfirst}6.125 & \cellcolor{tabfirst}17.380 
 & \cellcolor{tabfirst}0.201 & \cellcolor{tabfirst}0.553 & \cellcolor{tabsecond}0.884 & \cellcolor{tabfirst}0.976 \\
\midrule
Ours (Syn)$^{\dagger}$ & 0.422 & 0.776 & 0.899 & 0.927 & 0.891 & 6.315 & 19.703 & 30.739 & 0.891 & 0.968 & 0.994 & 0.996 \\
\bottomrule
\end{tabular}
\caption{\textbf{Performance Metrics for Template Prioritization.} The \colorbox{tabfirst}{best}, \colorbox{tabsecond}{second best}, and \colorbox{tabthird}{third best} results for the enzymatic dataset are color coded. The table is divided into three sets of metrics. The first set (columns 2-5) shows the accuracy of the presence of the literature rule in the top predictions (T1, T10, T50, T100). The second set (columns 6-9) represents the average number of applicable rules in the top predictions. The third set (columns 10-13) indicates the accuracy of the presence of at least one applicable rule in the top predictions.\\ $\dagger$ The synthetic model is trained and tested on a separate, larger dataset, while the other three models are trained and tested on the same dataset.}
\label{tab:model_metrics}
\end{table*}
To evaluate the effectiveness of our synthesis planning framework, we employed a comprehensive set of metrics designed to assess both the accuracy and applicability of the predicted reaction templates. Table \ref{tab:model_metrics} presents the performance of our model in comparison to existing state-of-the-art models, using three primary metrics:

\begin{enumerate}
    \item \textbf{Accuracy of the presence of the literature rule in the top n predictions (T1, T10, T50, T100):} This metric evaluates the model's ability to prioritize reaction templates documented in the literature. The "literature rule" refers to the reaction template extracted from the corresponding reaction in our dataset. This metric is crucial for assessing how well the model replicates known synthetic pathways. This is the commonly used metric in recent works \cite{levin2022merging, seidl2022improving}. While this metric is important for evaluating the replication of known synthetic pathways, assessing the model's ability to propose novel and feasible pathways is equally critical. Therefore, we also consider the following metrics, which provide a more comprehensive view of the model's performance. 

    \item \textbf{Average number of applicable rules in the top n predictions:} This metric provides insight into the diversity and feasibility of the proposed reaction templates. A rule is considered "applicable" when the RDKit library successfully applies the transformation to the target molecule, resulting in valid precursor structures. This metric is particularly important as it indicates the model's capability to explore broader chemical space and suggest alternative synthetic routes that may not be present in the literature but are chemically plausible. This is critical for our tree search, as a higher density of valid rules in the top predictions (T50, T100) expands the branching factor of the search tree with high-quality and newer candidates.

    \item \textbf{Accuracy of the presence of at least one applicable rule in the top n predictions:} This metric complements the other two by measuring the model's practical utility in retrosynthetic analysis. It assesses how often the model suggests at least one viable synthetic step, ensuring the progression of retrosynthetic analysis in chemically plausible directions. High scores here ensure the search algorithm rarely hits "dead ends".
\end{enumerate}

The introduction of these additional metrics (2 and 3) addresses the limitations of traditional evaluation methods, which often focus solely on predicting literature rules. While the accuracy of predicting literature rules is important, it does not fully capture the model's ability to suggest novel pathways. By incorporating metrics that consider rule applicability and diversity, we provide a more comprehensive evaluation that aligns with the exploratory nature of retrosynthetic planning.

Our enzymatic model (Ours (Enz)) demonstrates good performance across most metrics compared to the baseline DNN (Deep Neural Network) \cite{levin2022merging} and MHN \cite{seidl2022improving} models. It is important to contextualize the baseline performance, the DNN achieves a Top-1 accuracy of only 10\%, which underscores the inherent complexity of the retrosynthetic prediction task. Unlike standard classification problems, template prioritization involves selecting the correct chemical transformation from a massive search space of valid reaction templates. In this high-dimensional context, low absolute accuracy scores are standard, and incremental gains represent significant practical improvements.

First, compared to the standard DNN baseline \cite{levin2022merging}, our model delivers a large improvement. The DNN achieves a Top-1 literature accuracy of only 10\% and a Top-1 applicability presence of 12.8\%. Our model nearly doubles these figures to 18.3\% and 20.1\%, respectively. Given the high-dimensional search space (selecting 1 out of $>$17,000 templates), this jump represents a significant improvement in the model's ability to prioritize relevant chemistry. This trend is mirrored in the applicability metrics, indicating that our model not only retrieves ground truth better but also generates a higher number of valid chemical precursors.

Second, while our architecture shares similarities with the MHN model \cite{seidl2022improving}, our specific enhancements (Xavier initialization, rigorous dropout, and hyperparameter tuning) yield better convergence and robustness, particularly when predicting precursors for structurally complex or challenging target molecules. For example, in the \textit{Avg. Number of Applicable Rules} metric at T50, our model outperforms the MHN baseline (6.125 vs. 5.647). This indicates that our model retrieves approximately 8.5\% more valid chemical options in the top-50 candidates. This increased density of applicable rules provides the global greedy search algorithm with a richer pool of precursors, reducing the likelihood of missing a viable pathway.

Finally, the synthetic ensemble model (Ours (Syn)) demonstrates exceptional performance on the USPTO dataset, achieving a T1 accuracy of 42.2\%. Most notably, it exhibits a massive applicability rate: the Top-10 predictions yield, on average, 6.315 valid reaction rules. This high applicability ensures that the synthetic branch of our hybrid planner can almost always identify multiple feasible precursors, allowing the scoring system to aggressively optimize for cost and environmental impact without running out of chemical options.

It is important to note that the baseline DNN \cite{levin2022merging}, the MHN model \cite{seidl2022improving}, and our enzymatic model (Ours (Enz)) were trained and evaluated on the exact same enzymatic dataset splits, ensuring a direct and fair comparison.

\subsection*{3.2 \hspace{0.2cm} Comparison with pathways from the literature}

\begin{figure}[h!]

\begin{subfigure}{\textwidth}
    \centering
    \includegraphics[width=0.8\textwidth]{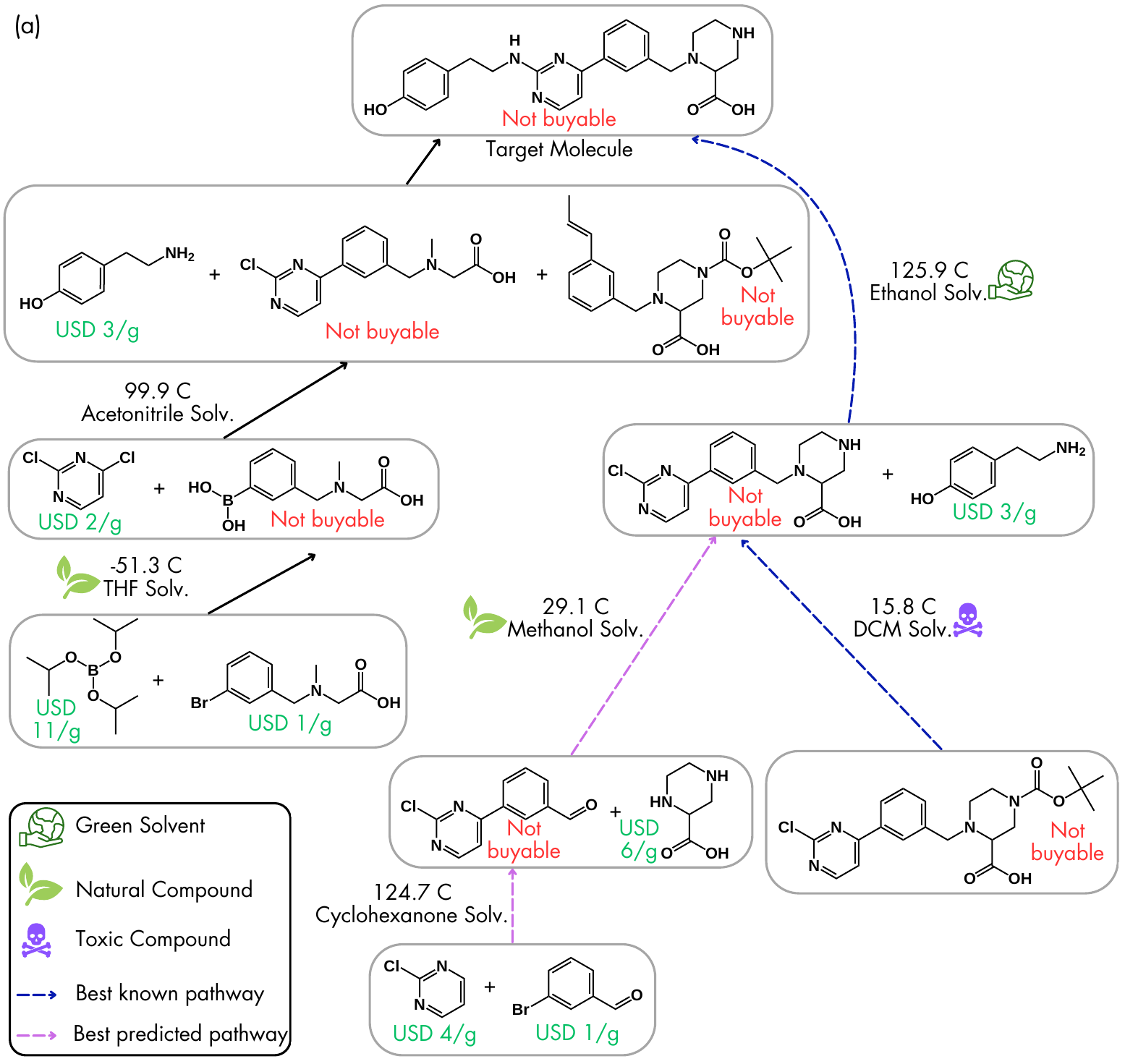}
    \label{fig:2a}
\end{subfigure}

\begin{subfigure}{\textwidth}
    \centering
    \includegraphics[width=\textwidth]{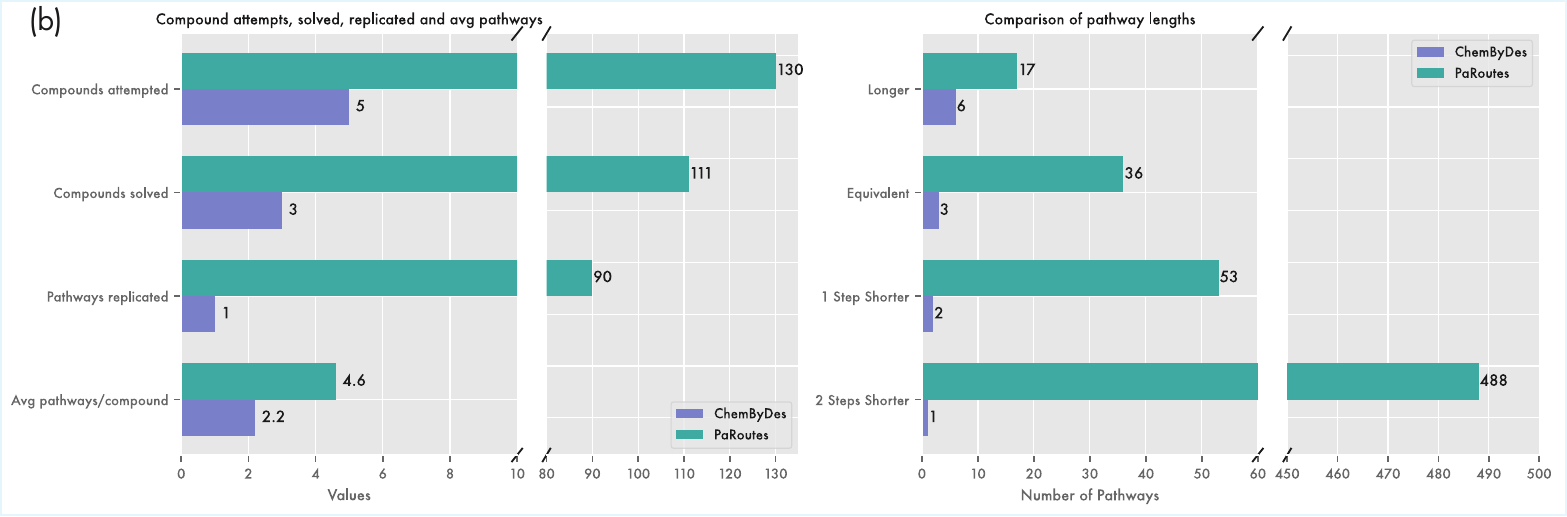}
    \label{fig:2b}
\end{subfigure}


\caption{\textbf{(a) Tree of reaction pathways.} The tree shows a representative example for a pathway presented in PaRoutes~\cite{D2DD00015F}. Our predicted pathway is producible using cheap precursors and less toxic, naturally occurring solvents. \\ \textbf{(b) Performance metrics for literature comparison.} These plots present the number of molecules solved, the average number of pathways predicted, and the distribution of predicted pathway lengths.}
\label{fig:lit_comp}    
\end{figure}

We assessed our synthesis planning framework by benchmarking it against established datasets and frameworks, including PaRoutes \cite{D2DD00015F} and ChemByDesign \cite{draghici2012chemistry}. Precisely, we assess the pathway lengths generated by our framework and report metrics on the number of replicated pathways and the average number of predicted pathways per molecule. We also present some case studies involving a hybrid approach combining enzymatic and synthetic pathways in \Cref{fig:lit_comp}(a) and Supplementary Figures 12 and 14. This comparative analysis aimed to assess our framework's accuracy, efficiency, and versatility in predicting viable synthetic routes for complex target molecules.

PaRoutes~\cite{D2DD00015F} is a robust benchmarking framework for evaluating multi-step retrosynthesis methods. It comprises two datasets of 10,000 synthetic routes derived from the patent literature alongside a curated list of purchasable molecules and reactions suitable for training retrosynthesis models. For this study, we utilized a scaffold split technique to select pathways associated with the ten most commonly occurring scaffolds in the PaRoutes dataset, which are given in  Supplementary Table 3. This approach enabled us to efficiently assess our framework's ability to predict synthetic routes that align with those documented in the patent literature while ensuring chemical diversity.

ChemByDesign~\cite{draghici2012chemistry}, on the other hand, is an online platform that organizes experimentally verified reaction pathways by name, year, and author. To ensure an unbiased evaluation, we focused on pathways discovered after 2021 that are under six steps long. These pathways were intentionally excluded from our training dataset to test the predictive capabilities of our framework on novel and unexplored synthetic routes. This strategy provided a unique opportunity to evaluate how effectively our framework generalizes to unseen data.

The results of this comparative analysis are summarized in Figure~\ref{fig:lit_comp}. \Cref{fig:lit_comp}(a) presents a representative tree of reaction pathways generated by our framework for a target molecule in the PaRoutes dataset. This evaluation was conducted on an expanded test set comprising 130 molecules from the PaRoutes dataset and 5 novel targets from ChemByDesign. Notably, our predicted pathway utilizes inexpensive precursors and environment-friendly solvents, such as ethanol and methanol, demonstrating its potential for sustainable synthesis planning. The tree also highlights alternative routes that leverage naturally occurring molecules while avoiding toxic solvents like dichloromethane (DCM); further examples of these green alternatives are illustrated in Supplementary Figure 9. These results underscore the framework's ability to prioritize green chemistry principles without compromising synthetic feasibility and cost.

\newpage

\newpage

\Cref{fig:lit_comp}(b) provides quantitative performance metrics for our framework compared to PaRoutes and ChemByDesign. The first bar plot illustrates the number of molecules attempted and successfully solved by each method. Our framework solved twelve out of fifteen molecules attempted. Furthermore, our framework replicated nine known pathways from the literature while suggesting a high number of pathways on average per molecule (4.8 and 2.2). These results demonstrate its ability to explore diverse chemical spaces and propose multiple viable options for synthetic planning.

Finally, the third bar plot compares the lengths of predicted pathways relative to known literature routes. Our framework identified thirty six shorter or equivalent-length pathways, with two cases$^\Omega$ yielding routes two steps shorter than those documented in PaRoutes and one case (Lupinine) yielding a route two steps shorter than that documented in ChemByDesign. This capability to optimize pathway length is particularly valuable in industrial settings where shorter synthetic routes can lead to significant cost savings and improved process efficiency. \{$\Omega$: N-(3-[5-Chloro-2-(difluoromethoxy)phenyl]-1-{2-[(4-pyridinylmethyl)amino]ethyl}-1H-pyrazol-4-yl)pyrazolo[1,5-a]pyrimidine-3-carboxamide and 4-(1-tert-butyl-4-oxo-5H-pyrazolo[4,3-c]pyridin-3-yl)thiophene-2-carboxamide\}

Supplementary Figure 9 illustrates an alternative pathway for Lupinine \cite{wang2024enantioselective} (found in ChemByDesign) synthesis predicted by MHNpath. Our framework identified a streamlined three-step route with moderate reaction conditions ($12.58\degree$C, $11.1\degree$C, and $-23.81\degree$C) and low precursor costs (\$87.36/g and \$0.10/g), contrasting with the five-step approach by Wang et al. that spans temperatures from $-78\degree$C to $85\degree$C. The test set of utilized from PaRoutes has been provided in the Supplementary Data.

\subsection*{3.3 \hspace{0.2cm} Comparison with other models}

\begin{figure}[H]

\begin{subfigure}{\textwidth}
    \centering
    \includegraphics[width=0.85\textwidth]{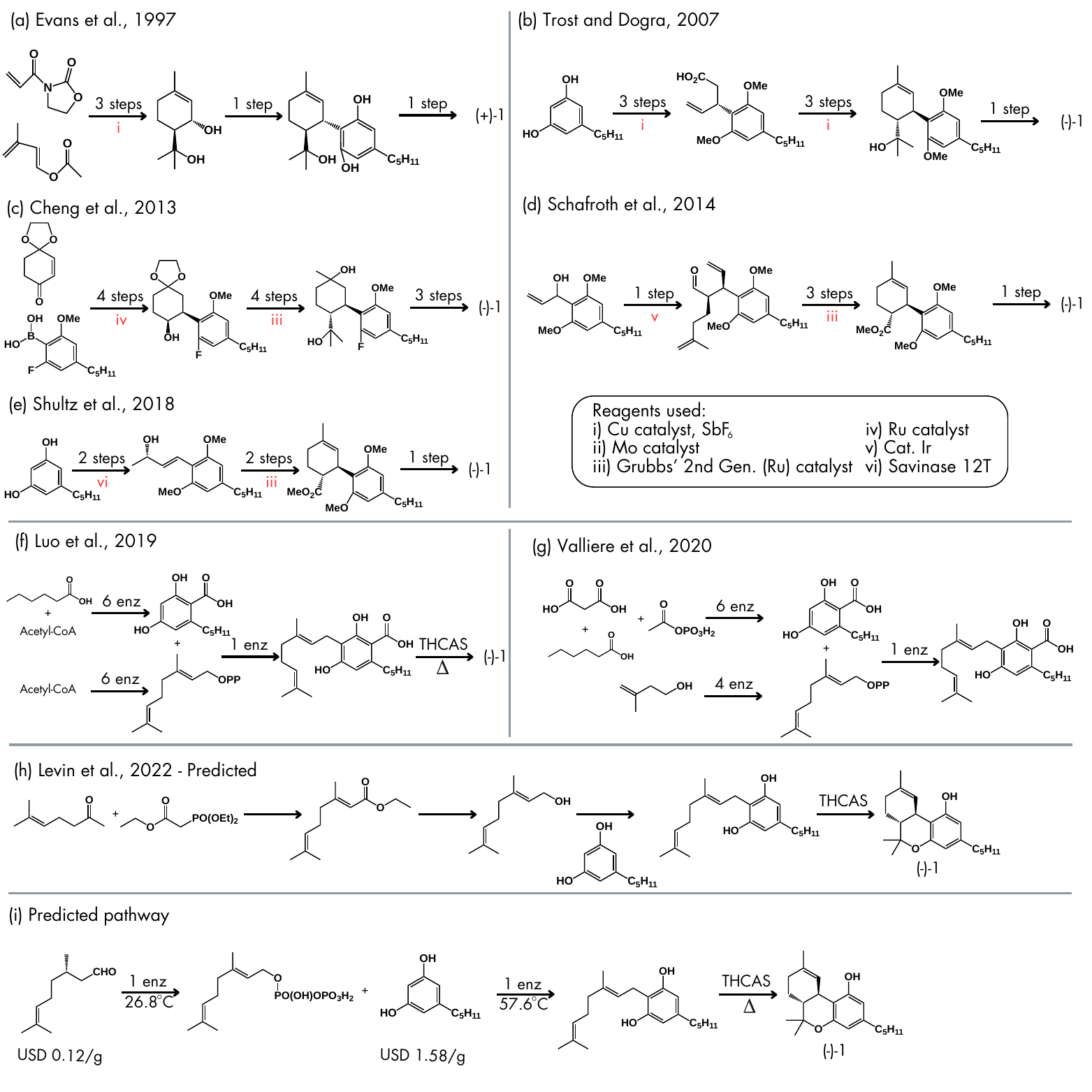}
    \label{fig:3a}
\end{subfigure}

\begin{subfigure}{\textwidth}
    \centering
    \includegraphics[width=0.99\textwidth]{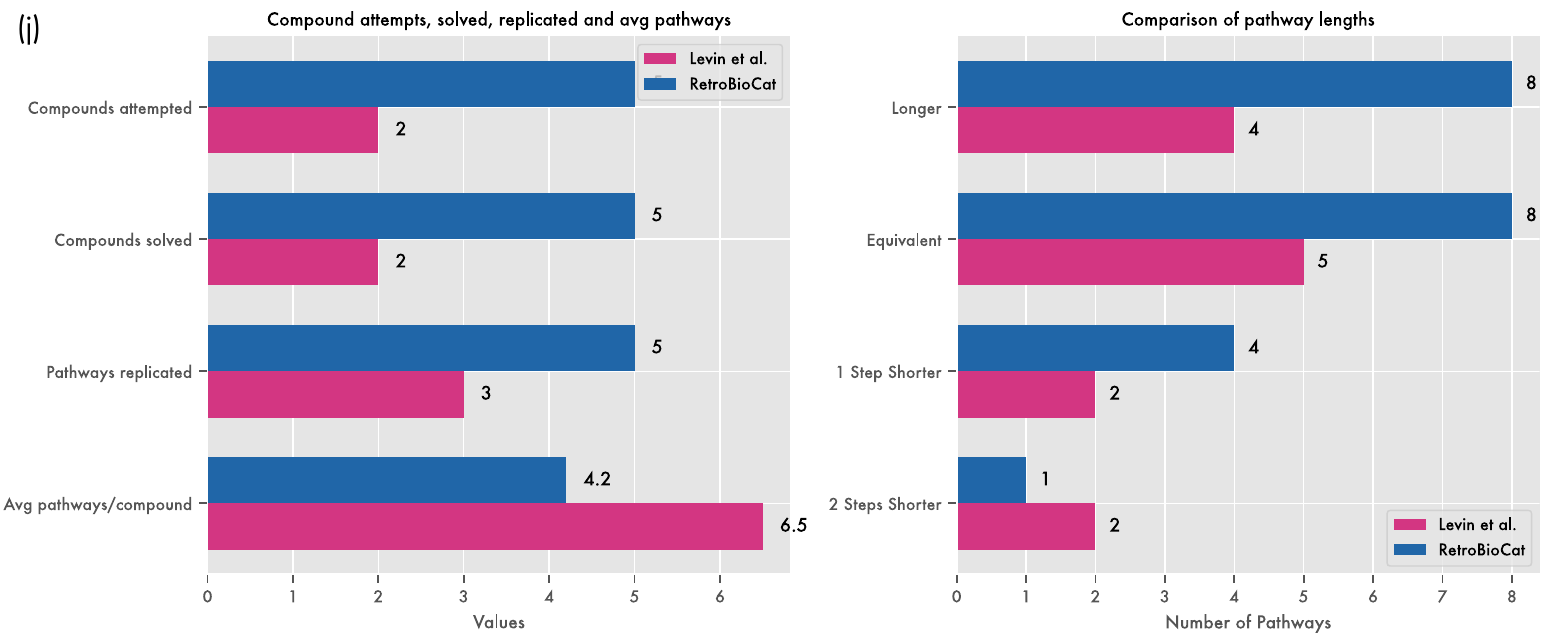}
    \label{fig:3b}
\end{subfigure}

\caption{\textbf{(a-g) Published pathways.} Overview of existing pathways to produce dronabinol. \textbf{(h) Previously predicted pathways by Levin et al. \cite{levin2022merging}.} Four-step reaction pathway to produce dronabinol as predicted by Levin et al. \cite{levin2022merging}. \textbf{(i) Our predicted pathway.} Three-step reaction to produce dronabinol from cheap precursors in ambient temperatures. We also replicated some of the other pathways. \textbf{(j) Performance metrics for comparison with other models.} These plots present the number of molecules solved, the average number of pathways predicted, and the distribution of predicted pathway lengths~\cite{levin2022merging, finnigan2021retrobiocat}.}
\label{fig:model_comp}
        
\end{figure}

We benchmark the performance of \name\ against two retrosynthetic planning tools: RetroBioCat \cite{finnigan2021retrobiocat} and the hybrid enzymatic-synthetic planner developed by Levin et al. \cite{levin2022merging}, as illustrated in \Cref{fig:model_comp}. Precisely, we assess the pathway lengths generated by our framework and report metrics on the number of replicated pathways and the average number of predicted pathways per compound. We also present some case studies involving a hybrid approach combining enzymatic and synthetic pathways in \Cref{fig:model_comp}(i) and Supplementary Figure 6. These comparisons highlight the advantages of our framework in terms of pathway length, cost-effectiveness, and the ability to replicate and improve upon previously reported pathways.

Although synthetic planning tools such as AiZynthFinder \cite{genheden2020aizynthfinder}, ASKCOS \cite{coley2019robotic}, Retro* \cite{chen2020retro}, and MHNreact \cite{seidl2022improving} represent significant benchmarks in the field, this section prioritizes comparisons with hybrid planners. The predictive capabilities of the underlying MLP and MHN-based engines used by these synthetic tools have already been quantitatively evaluated in Section 3.1 (Table \ref{tab:model_metrics}). We exclude a full tree search comparison with these platforms as they are restricted to purely synthetic routes and lack the integrated multi-objective scoring criteria central to \name, specifically real-time cost, toxicity, and temperature optimization. Furthermore, we do not report search time metrics because our framework relies on live API calls to retrieve dynamic pricing information. Since the total search duration is dominated by network latency rather than algorithmic efficiency, direct runtime comparisons with tools utilizing static building block datasets would be unrepresentative.

RetroBioCat \cite{finnigan2021retrobiocat} is a widely used platform for designing biocatalytic cascades. RetroBioCat facilitates the construction of selective and efficient biocatalytic pathways by leveraging an expanding enzyme toolbox and encoded reaction rules. Its strength lies in its ability to identify promising enzyme-specific routes, validated through several literature examples. However, as shown in \Cref{fig:model_comp} (i), our framework finds shorter pathways than RetroBioCat for many molecules like 4-ethenyl-2-fluorophenol and 2-phenylpiperidine. Specifically, \name\ successfully solved all five molecules attempted using RetroBioCat data, with an average of 4.2 pathways per molecule. The molecules attempted are listed in Supplementary Note 2. Moreover, our framework excelled in uncovering multiple shorter pathways and various pathways of equivalent length, offering users a range of options encompassing different enzymes, starting molecules, and temperature conditions.

Supplementary Figure 6 compares the pathways generated by RetroBioCat and \name. In the top panel, while the RetroBioCat pathway for a fluorinated compound involves multiple enzymatic steps (TPL, TAL, and DC), our approach achieves the same transformation using a single enzymatic step with TPT at 99$\degree$C, thereby significantly reducing both the complexity and the precursor cost to \$6.03/g. In the bottom panel, for an amine compound, the RetroBioCat route requires a three-step enzymatic cascade (CAR, TA, and IRED) that depends on cofactors such as NADPH and ATP. In contrast, our hybrid pathway utilizes only two enzymatic steps (PT at 54$\degree$C and PP at 10.1$\degree$C), with considerably lower precursor costs (\$0.1/g and \$1.8/g, respectively).

\newpage

\newpage

The hybrid enzymatic-synthetic planner by Levin et al. \cite{levin2022merging} represents another state-of-the-art approach that combines neural networks trained on extensive reaction databases with enzymatic transformations. Levin's model offers hybrid strategies for complex molecules such as THC and R,R-formoterol, showcasing the potential of enzyme-synthetic integration. However, our results demonstrate that \name\ not only replicates several of Levin's proposed pathways but also identifies novel alternatives that are shorter and more cost-effective.

As shown in \Cref{fig:model_comp} (h) and (i), Levin's predicted pathway for dronabinol involves a four-step synthesis. In contrast, our framework proposes a three-step pathway that reduces synthesis costs to \$0.12/g at ambient temperatures. This significant cost reduction is achieved by incorporating an optimized enzymatic step that eliminates the need for high-temperature reactions, underscoring the practical advantages of our approach. Although the final step of the predicted pathways remains unchanged, our framework excels in exploring diverse routes to the penultimate molecule, owing to its global-greedy tree search algorithm and scoring methodology. In addition to replicating published pathways from Levin et al., our framework identifies four shorter additional pathways.

Furthermore, Supplementary Figure 8 presents a novel four-step synthetic pathway for the synthesis of arformoterol predicted by our model. It starts from inexpensive precursors (\$3.11/g, \$1.30/g, \$0.51/g, and \$0.24/g). In contrast, Levin et al.'s approach involves a more complex five-step biocatalytic cascade that requires multiple enzymes and cofactors. The detailed annotations of reaction conditions in our pathway, such as temperature, solvent, and reagent specifics, demonstrate the practical applicability of our hybrid strategy.

A key advantage of \name\ is its ability to balance exploration and exploitation during retrosynthetic planning. This is evident from the higher average number of pathways generated per molecule (6.5 and 4.2), which provides chemists with a broader range of options for optimizing synthesis strategies based on reaction temperature, cost, or solvent toxicity.

\section*{4 \hspace{0.25cm} Conclusions}

The development and implementation of the \name\ tool represent progress over existing retrosynthetic frameworks in computer-aided synthesis planning. By leveraging machine learning, we have created a framework that can efficiently predict retrosynthetic pathways and facilitate the exploration of diverse chemical spaces. Our template prioritizer outperformed existing methods, as demonstrated by its higher accuracies across specific metrics and increased number of applicable rules. Importantly, our framework was able to replicate gold-standard pathways from PaRoutes as well as novel experimental reaction routes reported in ChemByDesign. In some instances, the model not only reproduced known synthesis routes but also identified alternative pathways that were shorter and more cost-effective. For example, our approach discovered a three-step synthetic route for dronabinol, reducing the synthesis cost to \$0.12/g as compared to a previously reported four-step pathway. 

\name\ underscores how ML can be utilized in reducing the manual workload and minimizing the trial-and-error strategies traditionally associated with chemical synthesis. Combined with a tree search-based strategy, the machine learning model allows for the automated and efficient prediction of ideal reaction pathways. The tool's user-tunable criteria allow researchers to prioritize different aspects of the synthesis process, such as cost, reaction temperature, and toxicity of solvents and reagents. This adaptability is particularly beneficial for experimental chemists who can tailor the tool to meet their research needs and constraints.

Despite its capabilities, \name\ has some limitations that outline future work directions. One potential issue is the reliance on predicted pathways without sufficient experimental validation, meaning that while the tool can suggest plausible synthetic routes, it cannot guarantee their success in practice and should be used as a starting point for further experimental investigation rather than a definitive solution. Additionally, the tool cannot effectively address enantiomer selectivity issues, as it does not inherently account for the stereoselective outcomes of reactions involving chiral molecules. Furthermore, \name\ is limited in its ability to provide detailed mechanistic insights into the predicted reactions, a drawback for users who require a deeper understanding of the underlying processes to optimize and troubleshoot reactions. Its overall accuracy and reliability are also heavily dependent on the quality and comprehensiveness of the underlying datasets; incomplete or biased data can lead to inaccurate predictions and missed opportunities for novel synthetic routes. To address these challenges, future work will focus on integrating enantioselective predictions by incorporating enantiomer-specific data and developing algorithms capable of predicting stereoselective outcomes, providing more mechanistic information through the integration of mechanistic databases, and continuously expanding and diversifying the datasets via collaborations and feasibility tests with experimental chemists.

In summary, \name\ demonstrates the potential of ML-driven tools in advancing CASP. Our results highlight this promise: for example, the enzymatic model achieved a top-1 accuracy of 18.3\%, a marked improvement over the 10\% of the DNN baseline and comparable to the 18.1\% of MHN. On the other hand, the synthetic ensemble model reached impressive 42.2\% top-1 accuracy. Moreover, our framework delivered an average of approximately 6 applicable rules in the top 10 predictions, highlighting its ability to explore diverse chemical spaces. In comparisons with pathways in the literature, \name\ successfully solved 12 out of 15 molecules attempted and even identified a three-step synthetic route for dronabinol that reduces costs to \$0.12/g at ambient temperatures, outperforming a competing four-step pathway. However, addressing its limitations and mitigating potential misuse remains essential to ensure the long-term success and reliability \name for assisting in synthesis of complex organic molecules.

\section*{Data Availability}

The synthetic dataset utilized in this study was obtained from the USPTO \cite{Lowe2017} repository, while the enzymatic dataset was sourced from the BKMS \cite{lang2011bkm} and RHEA \cite{bansal2022rhea} databases. All datasets are publicly accessible and open-source. The processed datasets generated and analyzed during the study and the trained model weights are available on \href{https://figshare.com/articles/dataset/Training_data_trained_models_and_other_required_files_for_A_User-Tunable_Machine_Learning_Framework_for_Step-Wise_Synthesis_Planning_/28673540}{Figshare}.
\section*{Code Availability}

The code used for data processing, model training, inference and the instructions to run our framework are available at \href{https://github.com/MSRG/mhnpath}{https://github.com/MSRG/mhnpath}.



\section*{Acknowledgement}

This research has received funding from the research project: ``Quantum Software Consortium: Exploring Distributed Quantum Solutions for Canada'' (QSC). QSC is financed under the National Sciences and Engineering Research Council of Canada (NSERC) Alliance Consortia Quantum Grants \#ALLRP587590-23. The authors also acknowledge the funding received from the University of Toronto's Data Sciences Institute (DSI) via its Catalyst Grant as well as the Canada Research Coordinating Committee's (CRCC) New Frontiers in Research Fund (NFRF) for their continued support. S. P. and V. K. P. thank DSI for providing financial support during the summer of 2025 via the Summer Undergraduate Data Science Research Opportunities Program and the Postdoctoral Fellowship, respectively. S. P. would also like to thank MolPort, Mcule and ChemSpace for providing API keys and access to a cost related database of chemical compounds. V.K.P. is grateful for the computational resource support provided by the Digital Research Alliance of Canada.

\newpage
\bibliography{references}

@article{finnigan2021retrobiocat,
  title={RetroBioCat as a computer-aided synthesis planning tool for biocatalytic reactions and cascades},
  author={Finnigan, William and Hepworth, Lorna J and Flitsch, Sabine L and Turner, Nicholas J},
  journal={Nature catalysis},
  volume={4},
  number={2},
  pages={98--104},
  year={2021},
  publisher={Nature Publishing Group UK London}
}

@article{sterling2015zinc,
  author = {Teague Sterling and John J. Irwin},
  title = {ZINC 15 – Ligand Discovery for Everyone},
  journal = {Journal of Chemical Information and Modeling},
  year = {2015},
  volume = {55},
  number = {11},
  pages = {2324--2337},
  month = nov,
  publisher = {American Chemical Society},
  doi = {10.1021/acs.jcim.5b00559},
  issn = {1549-9596},
  url = {https://doi.org/10.1021/acs.jcim.5b00559},
}

@article{razzaq2010continuous,
author = {Razzaq, Tahseen and Kappe, C.Oliver},
title = {Continuous Flow Organic Synthesis under High-Temperature/Pressure Conditions},
journal = {Chemistry – An Asian Journal},
volume = {5},
number = {6},
pages = {1274-1289},
doi = {https://doi.org/10.1002/asia.201000010},
url = {https://aces.onlinelibrary.wiley.com/doi/abs/10.1002/asia.201000010},
eprint = {https://aces.onlinelibrary.wiley.com/doi/pdf/10.1002/asia.201000010},
year = {2010}
}

@article{lang2011bkm,
  title={BKM-react, an integrated biochemical reaction database},
  author={Lang, Maren and Stelzer, Michael and Schomburg, Dietmar},
  journal={BMC biochemistry},
  volume={12},
  pages={1--9},
  year={2011},
  publisher={Springer}
}

@article{bansal2022rhea,
    author = {Bansal, Parit and Morgat, Anne and Axelsen, Kristian B and Muthukrishnan, Venkatesh and Coudert, Elisabeth and Aimo, Lucila and Hyka-Nouspikel, Nevila and Gasteiger, Elisabeth and Kerhornou, Arnaud and Neto, Teresa Batista and Pozzato, Monica and Blatter, Marie-Claude and Ignatchenko, Alex and Redaschi, Nicole and Bridge, Alan},
    title = "{Rhea, the reaction knowledgebase in 2022}",
    journal = {Nucleic Acids Research},
    volume = {50},
    number = {D1},
    pages = {D693-D700},
    year = {2021},
    month = {11},
    issn = {0305-1048},
    doi = {10.1093/nar/gkab1016},
    url = {https://doi.org/10.1093/nar/gkab1016},
    eprint = {https://academic.oup.com/nar/article-pdf/50/D1/D693/42058388/gkab1016.pdf},
}

@article{Landrum2016RDKit2016_09_4,
  author = {Landrum, Greg},
  description = {Release 2016_09_4 (Q3 2016) Release · rdkit/rdkit},
  title = {RDKit: Open-Source Cheminformatics Software},
  url = {https://github.com/rdkit/rdkit/releases
         /tag/Release\_2016\_09\_4},
  year = 2016
}

@article{Lowe2017,
author = "Daniel Lowe",
title = "{Chemical reactions from US patents (1976-Sep2016)}",
year = "2017",
month = "6",
journal = "",
url = "https://figshare.com/articles/dataset/Chemical
\_reactions\_from\_US\_patents\_1976-Sep2016\_/5104873",
doi = "10.6084/m9.figshare.5104873.v1"
}

@article{ramsauer2020hopfield,
  title={Hopfield networks is all you need},
  author={Ramsauer, Hubert and Sch{\"a}fl, Bernhard and Lehner, Johannes and Seidl, Philipp and Widrich, Michael and Adler, Thomas and Gruber, Lukas and Holzleitner, Markus and Pavlovi{\'c}, Milena and Sandve, Geir Kjetil and others},
  journal={arXiv preprint arXiv:2008.02217},
  year={2020}
}

@article{seidl2022improving,
  title={Improving few-and zero-shot reaction template prediction using modern hopfield networks},
  author={Seidl, Philipp and Renz, Philipp and Dyubankova, Natalia and Neves, Paulo and Verhoeven, Jonas and Wegner, Jorg K and Segler, Marwin and Hochreiter, Sepp and Klambauer, Gunter},
  journal={Journal of chemical information and modeling},
  volume={62},
  number={9},
  pages={2111--2120},
  year={2022},
  publisher={ACS Publications}
}

@article{coley2017prediction,
  author       = {Connor W. Coley and Regina Barzilay and Tommi S. Jaakkola and William H. Green and Klavs F. Jensen},
  title        = {Prediction of Organic Reaction Outcomes Using Machine Learning},
  journal      = {ACS Central Science},
  year         = {2017},
  volume       = {3},
  number       = {5},
  pages        = {434--443},
  publisher    = {American Chemical Society},
  doi          = {10.1021/acscentsci.7b00064},
  issn         = {2374-7943},
  url          = {https://doi.org/10.1021/acscentsci.7b00064}
}

@article{segler2017neural,
author = {Segler, Marwin H. S. and Waller, Mark P.},
title = {Neural-Symbolic Machine Learning for Retrosynthesis and Reaction Prediction},
journal = {Chemistry – A European Journal},
volume = {23},
number = {25},
pages = {5966-5971},
doi = {https://doi.org/10.1002/chem.201605499},
url = {https://chemistry-europe.onlinelibrary.wiley.com/doi/abs/10.1002/chem.201605499},
eprint = {https://chemistry-europe.onlinelibrary.wiley.com/doi/pdf/10.1002/chem.201605499},
year = {2017}
}

@article{coley2019robotic,
  title={A robotic platform for flow synthesis of organic compounds informed by AI planning},
  author={Coley, Connor W and Thomas III, Dale A and Lummiss, Justin AM and Jaworski, Jonathan N and Breen, Christopher P and Schultz, Victor and Hart, Travis and Fishman, Joshua S and Rogers, Luke and Gao, Hanyu and others},
  journal={Science},
  volume={365},
  number={6453},
  pages={eaax1566},
  year={2019},
  publisher={American Association for the Advancement of Science}
}

@article{gao2018using,
  title={Using machine learning to predict suitable conditions for organic reactions},
  author={Gao, Hanyu and Struble, Thomas J and Coley, Connor W and Wang, Yuran and Green, William H and Jensen, Klavs F},
  journal={ACS central science},
  volume={4},
  number={11},
  pages={1465--1476},
  year={2018},
  publisher={ACS Publications}
}

@article{levin2022merging,
  title={Merging enzymatic and synthetic chemistry with computational synthesis planning},
  author={Levin, Itai and Liu, Mengjie and Voigt, Christopher A and Coley, Connor W},
  journal={Nature Communications},
  volume={13},
  number={1},
  pages={7747},
  year={2022},
  publisher={Nature Publishing Group UK London}
}

@software{hadrien_mary_2024_10535844,
  author       = {Hadrien Mary and
                  Emmanuel Noutahi and
                  DomInvivo and
                  Lu Zhu and
                  Michel Moreau and
                  Steven Pak and
                  Desmond Gilmour and
                  Shawn Whitfield and
                  t and
                  Valence-JonnyHsu and
                  Honoré Hounwanou and
                  Ishan Kumar and
                  Saurav Maheshkar and
                  Shuya Nakata and
                  Kyle M. Kovary and
                  Cas Wognum and
                  Michael Craig and
                  DeepSource Bot},
  title        = {datamol-io/datamol: 0.12.3},
  month        = jan,
  year         = 2024,
  publisher    = {Zenodo},
  version      = {0.12.3},
  doi          = {10.5281/zenodo.10535844},
  url          = {https://doi.org/10.5281/zenodo.10535844}
}

@InProceedings{pmlr-v9-glorot10a,
  title = 	 {Understanding the difficulty of training deep feedforward neural networks},
  author = 	 {Glorot, Xavier and Bengio, Yoshua},
  booktitle = 	 {Proceedings of the Thirteenth International Conference on Artificial Intelligence and Statistics},
  pages = 	 {249--256},
  year = 	 {2010},
  editor = 	 {Teh, Yee Whye and Titterington, Mike},
  volume = 	 {9},
  series = 	 {Proceedings of Machine Learning Research},
  address = 	 {Chia Laguna Resort, Sardinia, Italy},
  month = 	 {13--15 May},
  publisher =    {PMLR},
  url = 	 {https://proceedings.mlr.press/v9/glorot10a.html}
}

@inproceedings{ioffe2015batch,
  title={Batch normalization: Accelerating deep network training by reducing internal covariate shift},
  author={Ioffe, Sergey and Szegedy, Christian},
  booktitle={International conference on machine learning},
  pages={448--456},
  year={2015},
  organization={pmlr}
}

@article{Hart1968,
  doi = {10.1109/tssc.1968.300136},
  url = {https://doi.org/10.1109/tssc.1968.300136},
  year = {1968},
  publisher = {Institute of Electrical and Electronics Engineers ({IEEE})},
  volume = {4},
  number = {2},
  pages = {100--107},
  author = {Peter Hart and Nils Nilsson and Bertram Raphael},
  title = {A Formal Basis for the Heuristic Determination of Minimum Cost Paths},
  journal = {{IEEE} Transactions on Systems Science and Cybernetics}
}

@article{Kiss2012,
    title = {http://Mcule.com: a public web service for drug discovery},
    author = {Kiss, Robert and Sandor, Mark and Szalai, Ferenc A},
    journal = {Journal of Cheminformatics},
    volume = {4},
    number = {Suppl 1},
    pages = {P17},
    year = {2012},
    publisher = {BioMed Central},
    address = {Budapest, H-1096, Hungary},
    doi = {10.1186/1758-2946-4-S1-P17},
    url = {https://doi.org/10.1186/1758-2946-4-S1-P17},
    issn = {1758-2946},
    pmcid = {PMC3341220},
    note = {Copyright 2012 Kiss et al; licensee BioMed Central Ltd.},
    language = {eng}
}

@misc{MolPort2020,
    title = {Easy compound ordering service - MolPort},
    author = {Molport},
    howpublished = {\url{https://www.molport.com/shop/index}}
}

@misc{Sorkun2024,
    author = {Sorkun, Mert C. and Saliou, Boris and Er, Sibel},
    title = {ChemPrice, a Python package for automated chemical price search},
    year = {2024},
    howpublished = {ChemRxiv},
    doi = {10.26434/chemrxiv-2024-1bxgg}
}

@misc{ACSGreenChemistryInstitute2011,
    author = {{ACS Green Chemistry Institute® Pharmaceutical Roundtable}},
    title = {Solvent Selection Guide: Version 2.0},
    year = {2011},
    month = {March 21},
    note = {Retrieved 12th May 2024 from: \url{http://www.acs.org/content/acs/en/greenchemistry/research-innovation/tools-for-green-chemistry.html}}
}

@article{10.1093/nar/gkac1008,
    author = {Gallo, Kathleen and Kemmler, Emanuel and Goede, Andrean and Becker, Finnja and Dunkel, Mathias and Preissner, Robert and Banerjee, Priyanka},
    title = "{SuperNatural 3.0—a database of natural products and natural product-based derivatives}",
    journal = {Nucleic Acids Research},
    volume = {51},
    number = {D1},
    pages = {D654-D659},
    year = {2022},
    month = {11},
    issn = {0305-1048},
    doi = {10.1093/nar/gkac1008},
    url = {https://doi.org/10.1093/nar/gkac1008},
    eprint = {https://academic.oup.com/nar/article-pdf/51/D1/D654/48440479/gkac1008.pdf},
}

@article{10.1093/nar/gku1004,
    author = {Wishart, David and Arndt, David and Pon, Allison and Sajed, Tanvir and Guo, An Chi and Djoumbou, Yannick and Knox, Craig and Wilson, Michael and Liang, Yongjie and Grant, Jason and Liu, Yifeng and Goldansaz, Seyed Ali and Rappaport, Stephen M.},
    title = "{T3DB: the toxic exposome database}",
    journal = {Nucleic Acids Research},
    volume = {43},
    number = {D1},
    pages = {D928-D934},
    year = {2014},
    month = {11},
    issn = {0305-1048},
    doi = {10.1093/nar/gku1004},
    url = {https://doi.org/10.1093/nar/gku1004},
    eprint = {https://academic.oup.com/nar/article-pdf/43/D1/D928/7311219/gku1004.pdf},
}

@article{szymkuc2016computer,
  title={Computer-aided synthesis design: 40 years on},
  author={Szymkuć, Szymon and Gajewska, Ewa P and Klucznik, Tomasz and Molga, Katarzyna and Dittwald, Piotr and Startek, Michał and Bajczyk, Marcin and Grzybowski, Bartosz A},
  journal={Angewandte Chemie International Edition},
  volume={55},
  number={20},
  pages={5904--5937},
  year={2016},
  publisher={Wiley Online Library}
}

@article{hart1968formal,
  title={A formal basis for the heuristic determination of minimum cost paths},
  author={Hart, Peter E and Nilsson, Nils J and Raphael, Bertram},
  journal={IEEE Transactions on Systems Science and Cybernetics},
  volume={4},
  number={2},
  pages={100--107},
  year={1968},
  publisher={IEEE}
}

@article{schwaller2019molecular,
  title={Molecular Transformer: A Model for Uncertainty-Calibrated Chemical Reaction Prediction},
  author={Schwaller, Philippe and Petraglia, Riccardo and Zullo, Valerio and Nair, Vishnu H. and Haeuselmann, Rico Andreas and Pisoni, Riccardo and Bekas, Costas and Laino, Teodoro},
  journal={Chemical Science},
  volume={10},
  number={9},
  pages={369--377},
  year={2019},
  publisher={Royal Society of Chemistry}
}

@Article{D1SC02362D,
author ="Kreutter, David and Schwaller, Philippe and Reymond, Jean-Louis",
title  ="Predicting enzymatic reactions with a molecular transformer",
journal  ="Chem. Sci.",
year  ="2021",
volume  ="12",
issue  ="25",
pages  ="8648-8659",
publisher  ="The Royal Society of Chemistry",
doi  ="10.1039/D1SC02362D",
url  ="http://dx.doi.org/10.1039/D1SC02362D"}

@article{tetko2020state,
  author       = {Igor V. Tetko and Pavel Karpov and Ruud Van Deursen and Guillaume Godin},
  title        = {State-of-the-art augmented NLP transformer models for direct and single-step retrosynthesis},
  journal      = {Nature Communications},
  year         = {2020},
  volume       = {11},
  number       = {1},
  pages        = {5575},
  doi          = {10.1038/s41467-020-19266-y},
  issn         = {2041-1723},
  url          = {https://doi.org/10.1038/s41467-020-19266-y}
}

@article{Irwin_2022,
doi = {10.1088/2632-2153/ac3ffb},
url = {https://dx.doi.org/10.1088/2632-2153/ac3ffb},
year = {2022},
month = {jan},
publisher = {IOP Publishing},
volume = {3},
number = {1},
pages = {015022},
author = {Irwin, Ross and Dimitriadis, Spyridon and He, Jiazhen and Bjerrum, Esben Jannik},
title = {Chemformer: a pre-trained transformer for computational chemistry},
journal = {Machine Learning: Science and Technology}
}

@article{WANG2021129845,
title = {RetroPrime: A Diverse, plausible and Transformer-based method for Single-Step retrosynthesis predictions},
journal = {Chemical Engineering Journal},
volume = {420},
pages = {129845},
year = {2021},
issn = {1385-8947},
doi = {https://doi.org/10.1016/j.cej.2021.129845},
url = {https://www.sciencedirect.com/science/article/pii/S1385894721014303},
author = {Xiaorui Wang and Yuquan Li and Jiezhong Qiu and Guangyong Chen and Huanxiang Liu and Benben Liao and Chang-Yu Hsieh and Xiaojun Yao}
}

@article{corey1985computer,
  title={Computer-assisted analysis in organic synthesis},
  author={Corey, EJ and Long, AK and Rubenstein, SD},
  journal={Science},
  volume={228},
  number={4698},
  pages={408--418},
  year={1985},
  doi={10.1126/science.3838594},
  pmid={3838594},
  issn={0036-8075},
  language={English},
  support={Research Support, Non-U.S. Gov't, U.S. Gov't, P.H.S.},
  publisher={American Association for the Advancement of Science},
}

@Article{D5DD00153F,
author ="Granqvist, Emma and Mercado, Rocío and Genheden, Samuel",
title  ="Retrosynformer: planning multi-step chemical synthesis routes via a decision transformer",
journal  ="Digital Discovery",
year  ="2025",
pages  ="-",
publisher  ="RSC",
doi  ="10.1039/D5DD00153F",
url  ="http://dx.doi.org/10.1039/D5DD00153F"}

@inbook{pytorch2019,
author = {Paszke, Adam and Gross, Sam and Massa, Francisco and Lerer, Adam and Bradbury, James and Chanan, Gregory and Killeen, Trevor and Lin, Zeming and Gimelshein, Natalia and Antiga, Luca and Desmaison, Alban and K\"{o}pf, Andreas and Yang, Edward and DeVito, Zach and Raison, Martin and Tejani, Alykhan and Chilamkurthy, Sasank and Steiner, Benoit and Fang, Lu and Bai, Junjie and Chintala, Soumith},
title = {PyTorch: an imperative style, high-performance deep learning library},
year = {2019},
publisher = {Curran Associates Inc.},
address = {Red Hook, NY, USA},
booktitle = {Proceedings of the 33rd International Conference on Neural Information Processing Systems},
articleno = {721},
numpages = {12}
}

@article{genheden2020aizynthfinder,
  author    = {Samuel Genheden and Amol Thakkar and Veronika Chadimov{\'a} and Jean-Louis Reymond and Ola Engkvist and Esben Bjerrum},
  title     = {AiZynthFinder: a fast, robust and flexible open-source software for retrosynthetic planning},
  journal   = {Journal of Cheminformatics},
  year      = {2020},
  volume    = {12},
  number    = {1},
  pages     = {70},
  doi       = {10.1186/s13321-020-00472-1},
  url       = {https://doi.org/10.1186/s13321-020-00472-1},
  issn      = {1758-2946}
}

@Article{D2DD00015F,
author ="Genheden, Samuel and Bjerrum, Esben",
title  ="PaRoutes: towards a framework for benchmarking retrosynthesis route predictions",
journal  ="Digital Discovery",
year  ="2022",
volume  ="1",
issue  ="4",
pages  ="527-539",
publisher  ="RSC",
doi  ="10.1039/D2DD00015F",
url  ="http://dx.doi.org/10.1039/D2DD00015F"}

@article{draghici2012chemistry,
  title={Chemistry By Design: A Web-Based Educational Flashcard for Exploring Synthetic Organic Chemistry},
  author={Draghici, Cristian and Njardarson, Jon T.},
  journal={Journal of Chemical Education},
  volume={89},
  number={8},
  pages={1080--1082},
  year={2012},
  publisher={American Chemical Society},
  doi={10.1021/ed2006423},
  url={https://doi.org/10.1021/ed2006423}
}

@misc{chemspace,
  author = {{ChemSpace}},
  title = {{ChemSpace}},
  howpublished = {\url{https://chem-space.com}},
  note = {Accessed: 2024-02-21}
}

@article{nicolaou2008molecules,
  title={Molecules that changed the world},
  author={Nicolaou, KC and Montagnon, T},
  journal={Clinical Pharmacology and Pharmacokinetics},
  volume={14},
  year={2008}
}

@article{brown2016analysis,
  title={Analysis of Past and Present Synthetic Methodologies on Medicinal Chemistry: Where Have All the New Reactions Gone?},
  author={Brown, Dean G. and Bostr{\"o}m, Jonas},
  journal={Journal of Medicinal Chemistry},
  volume={59},
  number={10},
  pages={4443--4458},
  doi={10.1021/acs.jmedchem.5b01409},
  pmid={26571338},
  publisher={ACS Publications},
  address={United States},
  year={2016},
  date={2015-12-01},
  journalid={9716531}
}

@Article{B713736M,
author ="Sheldon, Roger A.",
title  ="The E Factor: fifteen years on",
journal  ="Green Chem.",
year  ="2007",
volume  ="9",
issue  ="12",
pages  ="1273-1283",
publisher  ="The Royal Society of Chemistry",
doi  ="10.1039/B713736M",
url  ="http://dx.doi.org/10.1039/B713736M"}

@article{Horvath2007,
  author = {Horváth, István T. and Anastas, Paul T.},
  title = {Innovations and Green Chemistry},
  journal = {Chemical Reviews},
  volume = {107},
  number = {6},
  pages = {2169--2173},
  year = {2007},
  month = {6},
  publisher = {American Chemical Society},
  doi = {10.1021/cr078380v},
  url = {https://doi.org/10.1021/cr078380v},
  issn = {0009-2665}
}

@article{Szymkuc2016,
  author = {Szymkuć, Sara and Gajewska, Ewa P and Klucznik, Tomasz and Molga, Karol and Dittwald, Piotr and Startek, Michał and Bajczyk, Michał and Grzybowski, Bartosz A},
  title = {Computer-Assisted Synthetic Planning: The End of the Beginning},
  journal = {Angewandte Chemie International Edition},
  volume = {55},
  number = {20},
  pages = {5904--5937},
  year = {2016},
  month = {5},
  doi = {10.1002/anie.201506101},
  publisher = {Wiley-VCH Verlag},
  language = {eng},
  issn = {1433-7851},
  pmid = {27062365}
}

@incollection{PENSAK1977,
  author = {PENSAK, DAVID A. and COREY, E. J.},
  title = {LHASA—Logic and Heuristics Applied to Synthetic Analysis},
  booktitle = {Computer-Assisted Organic Synthesis},
  series = {ACS Symposium Series},
  volume = {61},
  pages = {1--32},
  year = {1977},
  month = {6},
  publisher = {AMERICAN CHEMICAL SOCIETY},
  doi = {10.1021/bk-1977-0061.ch001},
  url = {https://doi.org/10.1021/bk-1977-0061.ch001},
  isbn = {9780841203945}
}

@inproceedings{chen2020retro,
  title={Retro*: Learning Retrosynthetic Planning with Neural Guided A* Search},
  author={Chen, Binghong and Li, Chengtao and Dai, Hanjun and Song, Le},
  booktitle={The 37th International Conference on Machine Learning (ICML 2020)},
  year={2020}
}

@Article{D0SC04184J,
author ="Wang, Xiaoxue and Qian, Yujie and Gao, Hanyu and Coley, Connor W. and Mo, Yiming and Barzilay, Regina and Jensen, Klavs F.",
title  ="Towards efficient discovery of green synthetic pathways with Monte Carlo tree search and reinforcement learning",
journal  ="Chem. Sci.",
year  ="2020",
volume  ="11",
issue  ="40",
pages  ="10959-10972",
publisher  ="The Royal Society of Chemistry",
doi  ="10.1039/D0SC04184J",
url  ="http://dx.doi.org/10.1039/D0SC04184J"}

@misc{yu2024doubleendedsynthesisplanninggoalconstrained,
      title={Double-Ended Synthesis Planning with Goal-Constrained Bidirectional Search}, 
      author={Kevin Yu and Jihye Roh and Ziang Li and Wenhao Gao and Runzhong Wang and Connor W. Coley},
      year={2024},
      eprint={2407.06334},
      archivePrefix={arXiv},
      primaryClass={cs.AI},
      url={https://arxiv.org/abs/2407.06334}, 
}

@Article{D4DD00120F,
author ="Avila, Claudio and West, Adam and Vicini, Anna C. and Waddington, William and Brearley, Christopher and Clarke, James and Derrick, Andrew M.",
title  ="Chemistry in a graph: modern insights into commercial organic synthesis planning",
journal  ="Digital Discovery",
year  ="2024",
volume  ="3",
issue  ="9",
pages  ="1682-1694",
publisher  ="RSC",
doi  ="10.1039/D4DD00120F",
url  ="http://dx.doi.org/10.1039/D4DD00120F"}

@article{wang2024enantioselective,
  title={Enantioselective Synthesis of the 1, 3-Dienyl-5-Alkyl-6-Oxy Motif: Method Development and Total Synthesis},
  author={Wang, Jie and Guo, Chuning and Liu, Yaqian and Ji, Yunpeng and Jia, Hongli and Li, Houhua},
  journal={Angewandte Chemie International Edition},
  volume={63},
  number={15},
  pages={e202400478},
  year={2024},
  publisher={Wiley Online Library}
}

@inproceedings{bradshaw2018a,
title={A Generative Model For Electron Paths},
author={John Bradshaw and Matt J. Kusner and Brooks Paige and Marwin H. S. Segler and José Miguel Hernández-Lobato},
booktitle={International Conference on Learning Representations},
year={2019},
url={https://openreview.net/forum?id=r1x4BnCqKX},
}

@article{sacha2021molecule,
  author       = {Mikołaj Sacha and Mikołaj Błaż and Piotr Byrski and Paweł Dąbrowski-Tumański and Mikołaj Chromiński and Rafał Loska and Paweł Włodarczyk-Pruszyński and Stanisław Jastrzębski},
  title        = {Molecule Edit Graph Attention Network: Modeling Chemical Reactions as Sequences of Graph Edits},
  journal      = {Journal of Chemical Information and Modeling},
  year         = {2021},
  volume       = {61},
  number       = {7},
  pages        = {3273--3284},
  publisher    = {American Chemical Society},
  doi          = {10.1021/acs.jcim.1c00537},
  issn         = {1549-9596},
  url          = {https://doi.org/10.1021/acs.jcim.1c00537}
}

@Article{C8SC04228D,
author ="Coley, Connor W. and Jin, Wengong and Rogers, Luke and Jamison, Timothy F. and Jaakkola, Tommi S. and Green, William H. and Barzilay, Regina and Jensen, Klavs F.",
title  ="A graph-convolutional neural network model for the prediction of chemical reactivity",
journal  ="Chem. Sci.",
year  ="2019",
volume  ="10",
issue  ="2",
pages  ="370-377",
publisher  ="The Royal Society of Chemistry",
doi  ="10.1039/C8SC04228D",
url  ="http://dx.doi.org/10.1039/C8SC04228D"}

@article{qian2020integrating,
  title={Integrating deep neural networks and symbolic inference for organic reactivity prediction},
  author={Qian, Wesley Wei and Russell, Nathan T and Simons, Claire LW and Luo, Yunan and Burke, Martin D and Peng, Jian},
  year={2020}
}

@InProceedings{pmlr-v139-bi21a,
  title = 	 {Non-Autoregressive Electron Redistribution Modeling for Reaction Prediction},
  author =       {Bi, Hangrui and Wang, Hengyi and Shi, Chence and Coley, Connor and Tang, Jian and Guo, Hongyu},
  booktitle = 	 {Proceedings of the 38th International Conference on Machine Learning},
  pages = 	 {904--913},
  year = 	 {2021},
  editor = 	 {Meila, Marina and Zhang, Tong},
  volume = 	 {139},
  series = 	 {Proceedings of Machine Learning Research},
  month = 	 {18--24 Jul},
  publisher =    {PMLR},
  url = 	 {https://proceedings.mlr.press/v139/bi21a.html}
}

@article{chen2022generalized,
  author       = {Shuan Chen and Yousung Jung},
  title        = {A generalized-template-based graph neural network for accurate organic reactivity prediction},
  journal      = {Nature Machine Intelligence},
  year         = {2022},
  volume       = {4},
  number       = {9},
  pages        = {772--780},
  doi          = {10.1038/s42256-022-00526-z},
  issn         = {2522-5839},
  url          = {https://doi.org/10.1038/s42256-022-00526-z}
}

@misc{figshare,
    title = {Figshare Data},
    author = {Shivesh Prakash, Viki Kumar Prasad, Hans-Arno Jacobsen},
    howpublished = {\url{https://figshare.com/articles/dataset/Training_data_trained_models_and_other_required_files_for_A_User-Tunable_Machine_Learning_Framework_for_Step-Wise_Synthesis_Planning_/28673540}}
}

@article{zhang2025groupretro,
  title={A data-driven group retrosynthesis planning model inspired by neurosymbolic programming},
  author={Zhang, Xuefeng and Lin, Haowei and Zhang, Muhan and Zhou, Yixin and Ma, Jun},
  journal={Nature Communications},
  volume={16},
  number={1},
  year={2025},
  doi={10.1038/s41467-024-55374-9}
}

@article{roh2025higherlevel,
  title={Higher-level Strategies for Computer-Aided Retrosynthesis},
  author={Roh, Jihye and Joung, Joonyoung F. and Yu, Kevin and Tu, Zhilei and Bartholomew, G. L. and Santiago-Reyes, O. A. and Fong, M. H. and Sarpong, Richmond and Reisman, Sarah E. and Coley, Connor W.},
  journal={ChemRxiv},
  year={2025},
  doi={10.26434/chemrxiv-2025-21zvt-v2}
}

@article{choe2025crosstalk,
  title={Retrosynthetic crosstalk between single-step reaction and multi-step planning},
  author={Choe, JunSeok and Kim, Hajung and Chok, Yan Ting and Gim, Minji and Kang, Joonghyuk},
  journal={Journal of Cheminformatics},
  volume={17},
  number={1},
  year={2025},
  doi={10.1186/s13321-025-01088-z}
}

@article{andronov2025specbeam,
  title={Fast and scalable retrosynthetic planning with a transformer neural network and speculative beam search},
  author={Andronov, Mikhail and Andronova, Natalia and Wand, Michael and Schmidhuber, J{\"u}rgen and Clevert, Djork-Arn{\'e}},
  journal={arXiv},
  volume={abs/2508.01459},
  year={2025},
  doi={10.48550/arxiv.2508.01459}
}

\newpage

\title{A User-Tunable Machine Learning Framework for Step-Wise Synthesis Planning: \\ Supplementary Information}
\author{
 {\normalfont $^1$Department of Computer Science, University of Toronto, 40 St George St, Toronto, ON M5S 2E4} \\
 {\normalfont $^2$The Edward S. Rogers Sr. Department of Electrical \& Computer Engineering, University of Toronto, 10 King's College Rd, Toronto, ON M5S 3G8} \\
 {\normalfont $^3$Data Science Institute, University of Toronto, 700 University Ave 10th floor, Toronto, ON M7A 2S4} \\
 {\normalfont $^4$Current Affiliation: Department of Chemistry, University of Calgary, 2500 University Drive NW, Calgary, AB T2N 1N4} \\
 $^{*}$\texttt{jacobsen@eecg.toronto.edu}
}
\date{}
\newcommand{\TAName}{Professor David Lindell}
\newcommand{\CourseName}{In submission to Nature Machine Intelligence}

\begin{titlepage}
    \centering
    \vspace*{2cm}
    {\huge A User-Tunable Machine Learning Framework for Step-Wise Synthesis Planning: \\ Supplementary Information \par}
    \vspace{1cm}
    {\Large Shivesh Prakash$^{1}$\href{https://orcid.org/0009-0007-4120-0921}{\includegraphics[width=10pt]{figures/ORCID.png}}\quad
    Nandan Patel$^{4}$\href{https://orcid.org/0009-0003-5660-5658}{\includegraphics[width=10pt]{figures/ORCID.png}}\quad
 Hans-Arno Jacobsen$^{1, 2}$\href{https://orcid.org/0000-0003-0813-0101}{\includegraphics[width=10pt]{figures/ORCID.png}}\quad 
 Viki Kumar Prasad$^{2, 3, 4 *}$\href{https://orcid.org/0000-0003-0982-3129}{\includegraphics[width=10pt]{figures/ORCID.png}} \par}
    {\normalfont {\normalfont $^1$Department of Computer Science, University of Toronto, 40 St George St, Toronto, ON M5S 2E4} \\
 {\normalfont $^2$The Edward S. Rogers Sr. Department of Electrical \& Computer Engineering, University of Toronto, 10 King's College Rd, Toronto, ON M5S 3G8} \\
 {\normalfont $^3$Data Science Institute, University of Toronto, 700 University Ave 10th floor, Toronto, ON M7A 2S4} \\
 {\normalfont $^4$Current affiliation: Department of Chemistry, University of Calgary, 2500 University Drive NW, Calgary, AB T2N 1N4} \\
 $^{*}$\texttt{vikikumar.prasad@ucalgary.ca} \par}
\end{titlepage}

\newpage

\tableofcontents

\newpage

\section{Supplementary Method 1: Hyperparameter tuning for enzymatic template prioritizer}

In this work, hyperparameter tuning was performed using a one‐factor‐at‐a‐time (OFAT) approach. Each experiment involved modifying a single hyperparameter from a baseline configuration while keeping all other settings constant. The performance of each model was assessed using an evaluation score, with lower validation loss indicating better performance.

Several aspects were considered during the tuning process:

\begin{itemize}
  \item \textbf{Training Epochs:} A comparison between 11 and 15 epochs revealed that extending the training duration did not improve performance; indeed, the model trained for 11 epochs performed better.
  \item \textbf{Concatenation Threshold:} Varying the concatenation threshold from 1 up to 6 showed progressive improvements until a threshold of 3 was reached, beyond which performance gains plateaued or reversed.
  \item \textbf{Dropout Rate:} Experiments with dropout values of 0, 0.01, and 0.1 suggested that a small dropout (0.01) was sufficient to mitigate overfitting without deteriorating the score.
  \item \textbf{Learning Rate:} A learning rate of \(1\times10^{-4}\) yielded more stable and improved convergence in comparison to a higher rate of \(2\times10^{-4}\).
  \item \textbf{Hopf Parameter Tuning:} Adjustments of the hopf beta parameter indicated that values around 0.035 resulted in superior performance. In addition, the adoption of the Tanh activation function for the hopf association further contributed to lowering the evaluation score.
  \item \textbf{Architectural Modifications:} Architectural parameters, including the number of layers in the template encoder and adjustments to batch size, were also evaluated. Setting the template encoder layers to 2 and using a batch size of 32 were identified as beneficial modifications.
\end{itemize}

Collectively, these experiments guided the selection of the final configuration, which integrated the optimal choices from each parameter sweep.

\newpage

\section{Supplementary Table 1: Experiments conducted for hyperparameter tuning of enzymatic template prioritizer}

Below is a summary of all experiments conducted, the specific value(s) tried for each hyperparameter, and the corresponding validation loss achieved.

\begin{table}[ht]
\centering
\begin{tabular}{@{}llcl@{}}
\toprule
\textbf{Experiment}    & \textbf{Hyperparameter}           & \textbf{Value}                   & \textbf{Validation Loss} \\ \midrule
Default (epoch 11)     & Epoch Count                       & 11                               & 6.077                    \\
Epoch 15               & Epoch Count                       & 15                               & 6.160                    \\ \midrule
concat 1               & Concat Threshold                  & 1                                & 5.754                    \\
concat 2               & Concat Threshold                  & 2                                & 5.505                    \\
concat 3               & Concat Threshold                  & 3                                & 5.472                    \\
concat 4               & Concat Threshold                  & 4                                & 5.551                    \\
concat 6               & Concat Threshold                  & 6                                & 5.542                    \\ \midrule
drop 1                 & Dropout Rate                      & 0.1                              & 5.465                    \\
drop 01                & Dropout Rate                      & 0.01                             & 5.462                    \\
drop 0                 & Dropout Rate                      & 0                                & 5.463                    \\ \midrule
lr 1                   & Learning Rate                     & \(1\times10^{-4}\)               & 5.276                    \\
lr 2                   & Learning Rate                     & \(2\times10^{-4}\)               & 5.361                    \\ \midrule
beta 02                & Hopf Beta                         & 0.02                             & 5.619                    \\
beta 2                 & Hopf Beta                         & 0.2                              & 6.338                    \\
beta 04                & Hopf Beta                         & 0.04                             & 5.199                    \\
beta 035               & Hopf Beta                         & 0.035                            & 5.194                    \\
beta 03                & Hopf Beta                         & 0.03                             & 5.234                    \\ \midrule
dim 1                  & Hopf Association Dimension        & 1024                             & 5.472                    \\ \midrule
tanh                   & Activation Function (Association) & Tanh                             & 5.194                    \\ \midrule
inpn                   & Input Normalization               & False                            & 5.156                    \\
ascn                   & Norm. (Input \& Association)       & False, False                     & 7.292                    \\ \midrule
lay2                   & Hopf Layers                       & 2                                & 5.178                    \\
lay3                   & Hopf Layers                       & 3                                & 5.193                    \\ \midrule
mol2                   & Mol Encoder Layers                & 2                                & 5.177                    \\ \midrule
tem2                   & Temp Encoder Layers               & 2                                & 5.033                    \\
tem3                   & Temp Encoder Layers               & 3                                & 5.085                    \\ \midrule
batch256               & Batch Size                        & 256                              & 5.068                    \\
batch1024              & Batch Size                        & 1024                             & 5.448                    \\
batch32                & Batch Size                        & 32                               & 5.051                    \\ \midrule
final 11               & Final Configuration               & (Optimal settings combined)      & 4.995                    \\ \bottomrule
\end{tabular}
\caption{Summary of hyperparameter tuning experiments, including the modified parameter, tested value, and achieved validation loss.}
\end{table}

\newpage

\section{Supplementary Method 2: Hyperparameter tuning for synthetic template prioritizers}

Due to constraints in both computational resources and the increased complexity of the synthetic template prioritizer models, we did not perform an extensive hyperparameter tuning process for this case. Instead, we adopted the optimal hyperparameters that were previously determined for the enzymatic template prioritizer. 

This decision was driven by several factors. First, the synthetic and enzymatic template prioritizers share a similar underlying architecture, meaning that the hyperparameters optimized for the enzymatic system are expected to generalize well to the synthetic case. Second, preliminary experiments indicated that applying these parameters to the synthetic model yielded robust and consistent performance, thereby reducing the necessity for additional tuning. Finally, redirecting computational resources away from redundant parameter searches allowed us to focus on further evaluations and ensure methodological consistency across both systems.

By leveraging the proven settings from the enzymatic template prioritizer, we maintained efficiency in our workflow while still achieving reliable performance for the synthetic template prioritizers.

\newpage

\section{Supplementary Table 2: Hyperparameters used for training enzymatic and synthetic template prioritizers}

\begin{table}[h]
\centering
\begin{tabular}{ll}
\toprule
\textbf{Parameter} & \textbf{Value} \\
\midrule
Concat Rand Template Threshold & 3 \\
Epochs & 11 \\
Dropout Rate & 0.01 \\
Learning Rate & \(1\times10^{-4}\) \\
Hopf Beta & 0.035 \\
Hopf Association Activation & Tanh \\
Normalization of Input & False \\
Template Encoder Layers & 2 \\
Batch Size & 32 \\
\bottomrule
\end{tabular}
\caption{Optimal hyperparameters for the enzymatic and synthetic template prioritizer.}
\end{table}

\newpage

\section{Supplementary Table 3: Top 10 most popular scaffolds from PaRoutes dataset \cite{D2DD00015F}}

To assess the structural diversity of our compound set, we performed a scaffold analysis using the Bemis-Murcko scaffold framework as implemented in RDKit \cite{Landrum2016RDKit2016_09_4}.

From the complete dataset of 20,000 molecules, we identified 11,801 unique scaffolds, indicating substantial structural diversity. A total of 19,742 molecules (98.7\%) yielded valid scaffolds, with the remaining compounds likely representing aliphatic chains or other structures without ring systems. Table \ref{tab:top_scaffolds} presents the ten most frequently occurring scaffolds in our dataset.

\begin{table}[htbp]
\centering
\caption{Top 10 molecular scaffolds identified in the dataset of 20,000 compounds.}
\label{tab:top_scaffolds}
\begin{tabular}{cccl}
\hline
\textbf{Rank} & \textbf{Count} & \textbf{Percentage (\%)} & \textbf{Scaffold (SMILES)} \\
\hline
1 & 1,209 & 6.12 & \texttt{c1ccccc1} \\
2 & 209 & 1.06 & \texttt{c1ccncc1} \\
3 & 144 & 0.73 & \texttt{c1ccc(-c2ccccc2)cc1} \\
4 & 118 & 0.60 & \texttt{c1ccc(COc2ccccc2)cc1} \\
5 & 95 & 0.48 & \texttt{c1ccc2[nH]ccc2c1} \\
6 & 85 & 0.43 & \texttt{c1ccc(Oc2ccccc2)cc1} \\
7 & 83 & 0.42 & \texttt{c1ccc2ncccc2c1} \\
8 & 65 & 0.33 & \texttt{C1CCCCC1} \\
9 & 61 & 0.31 & \texttt{c1ccc2ccccc2c1} \\
10 & 60 & 0.30 & \texttt{c1ccc(-c2ccccn2)cc1} \\
\hline
\end{tabular}
\end{table}

The most prevalent scaffold was benzene (\texttt{c1ccccc1}), appearing in 6.12\% of the compounds with valid scaffolds. Pyridine (\texttt{c1ccncc1}) was the second most common scaffold, representing 1.06\% of the dataset. 

\newpage

\section{Supplementary Figure 1: Accuracy metrics during training of enzymatic template prioritizer}

\begin{figure}[h!]
\begin{centering}
    \includegraphics[width=0.68\textwidth]{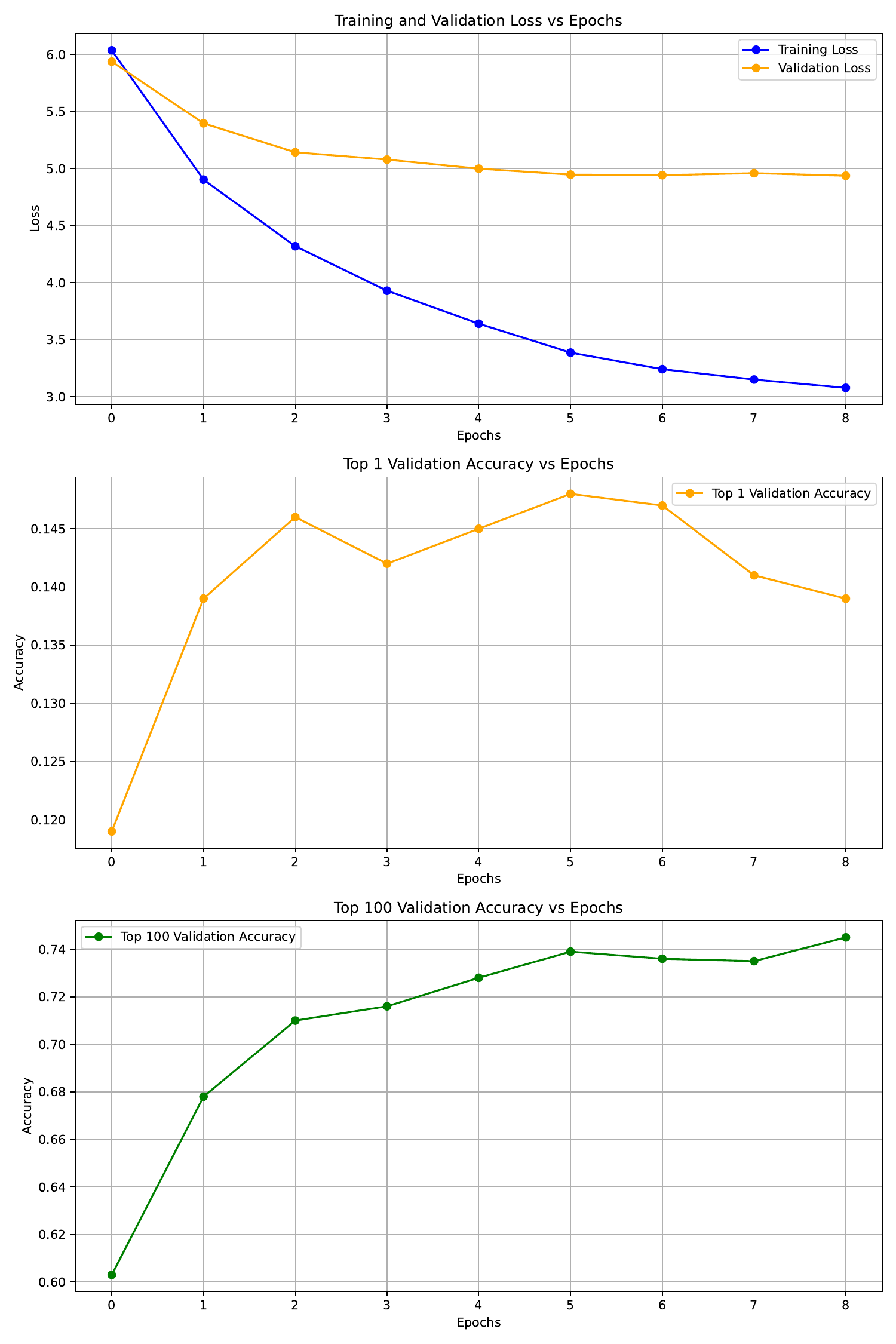}
    \caption{Training and validation metrics for the enzymatic template prioritizer over 10 epochs. The top panel shows the training and validation loss. The middle panel illustrates the top-1 validation accuracy. The bottom panel depicts the top-100 validation accuracy.}
\end{centering}
\end{figure}

\newpage

\section{Supplementary Figure 2: Accuracy metrics during training of the first synthetic template prioritizer}

\begin{figure}[h!]
\begin{centering}
    \includegraphics[width=0.68\textwidth]{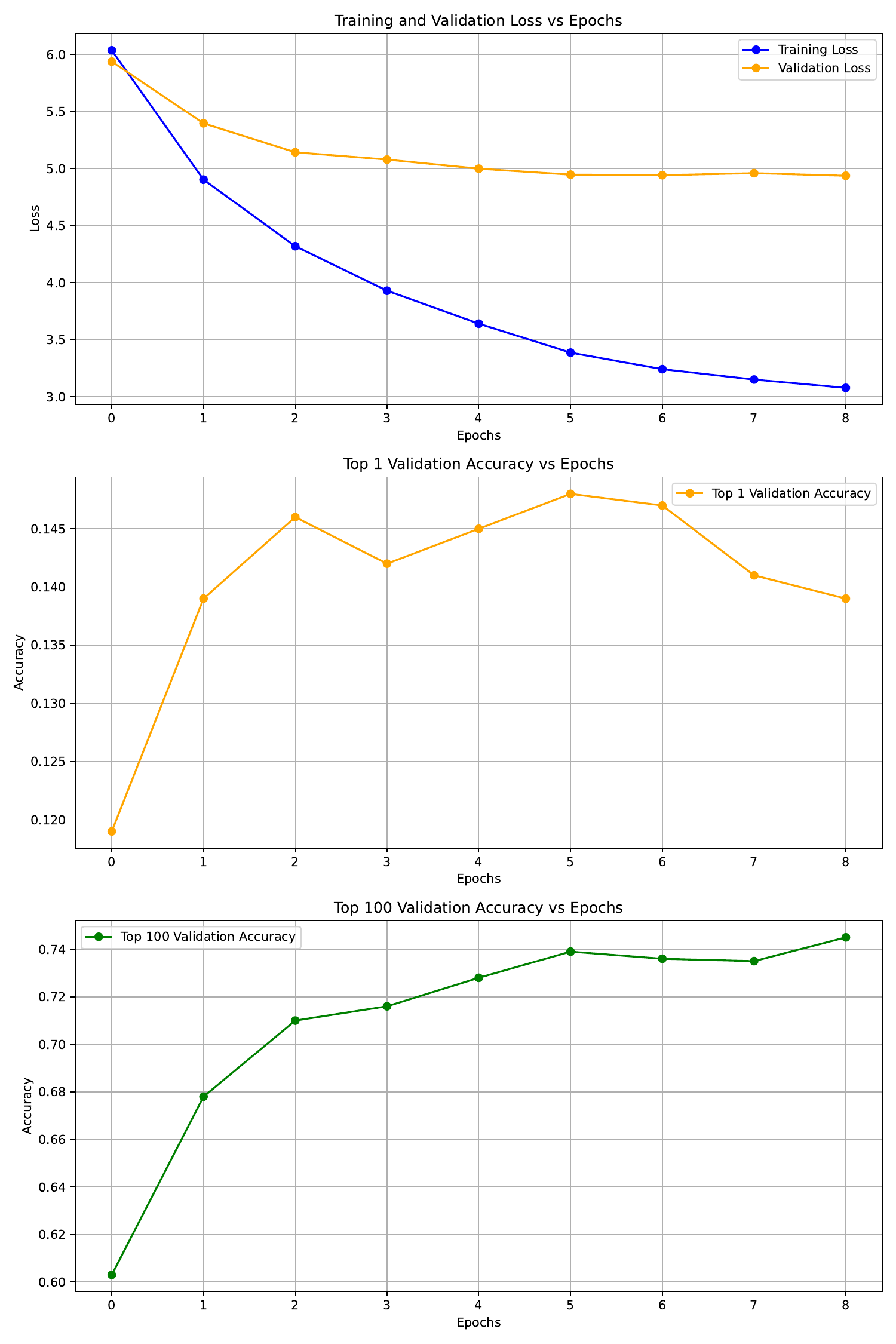}
    \caption{Training and validation metrics for the first synthetic template prioritizer over 10 epochs. The top panel shows the training and validation loss. The middle panel illustrates the top-1 validation accuracy. The bottom panel depicts the top-100 validation accuracy.}
\end{centering}
\end{figure}

\newpage

\section{Supplementary Figure 3: Synthetic pathway tree for 2-phenoxyethanamine}

\begin{figure}[h!]
\begin{centering}
    \includegraphics[width=\textwidth]{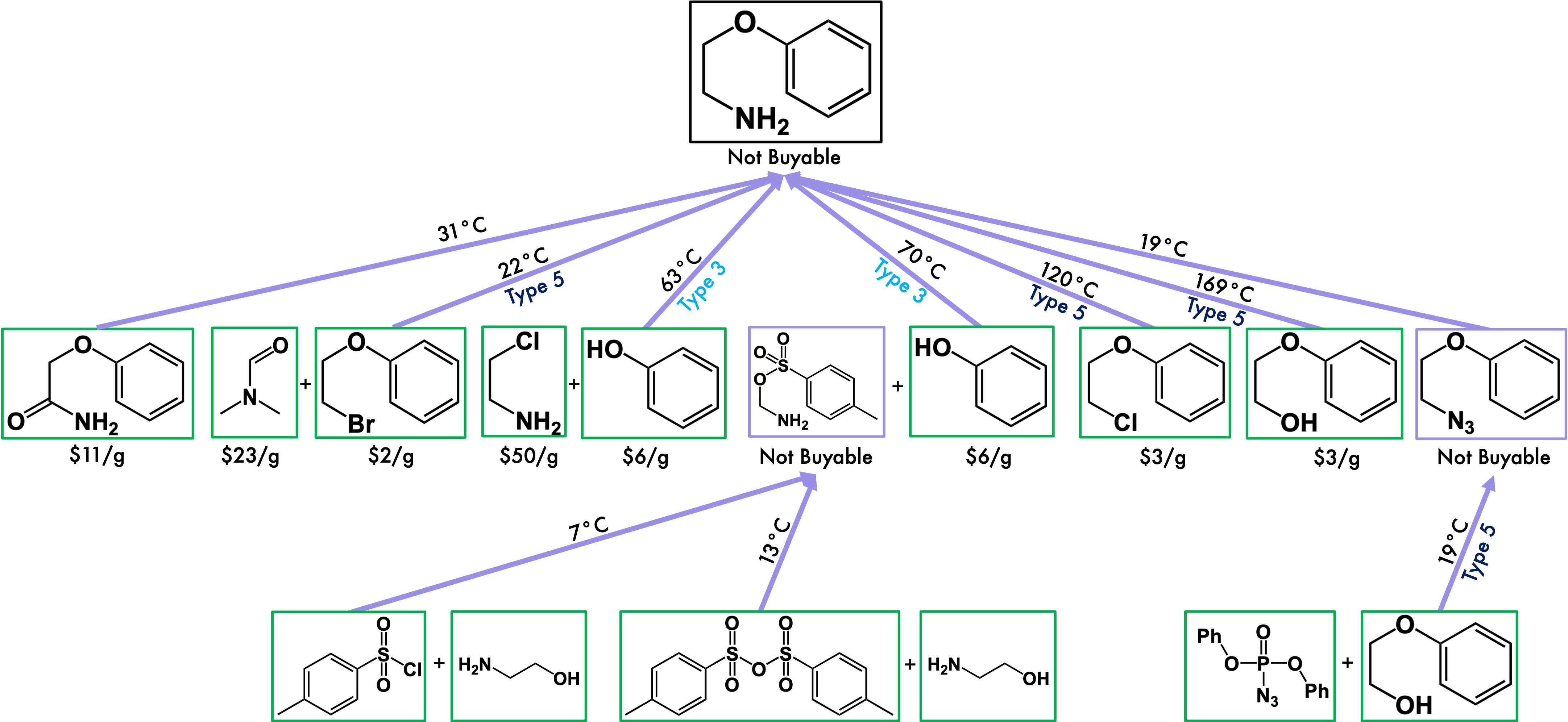}
    \caption{The synthetic pathway tree for 2-phenoxyethanamine illustrates potential reaction routes and intermediate compounds. The root node represents the target compound, marked as ``Not Buyable,'' indicating that it must be synthesized. Each branch corresponds to a reaction step, annotated with the reaction temperature and type. Intermediate compounds are shown with their respective costs in USD per gram. Green-bordered nodes represent commercially available compounds, while purple-bordered nodes denote intermediates that require further synthesis. This hierarchical structure highlights the complexity of the synthetic pathway and provides insight into cost-effective and feasible routes for the target compound.}
\end{centering}
\end{figure}

\newpage

\section{Supplementary Figure 4: Enzymatic, synthetic and hybrid pathways for 2-phenoxyethanamine}

\begin{figure}[h!]
\begin{centering}
    \includegraphics[width=\textwidth]{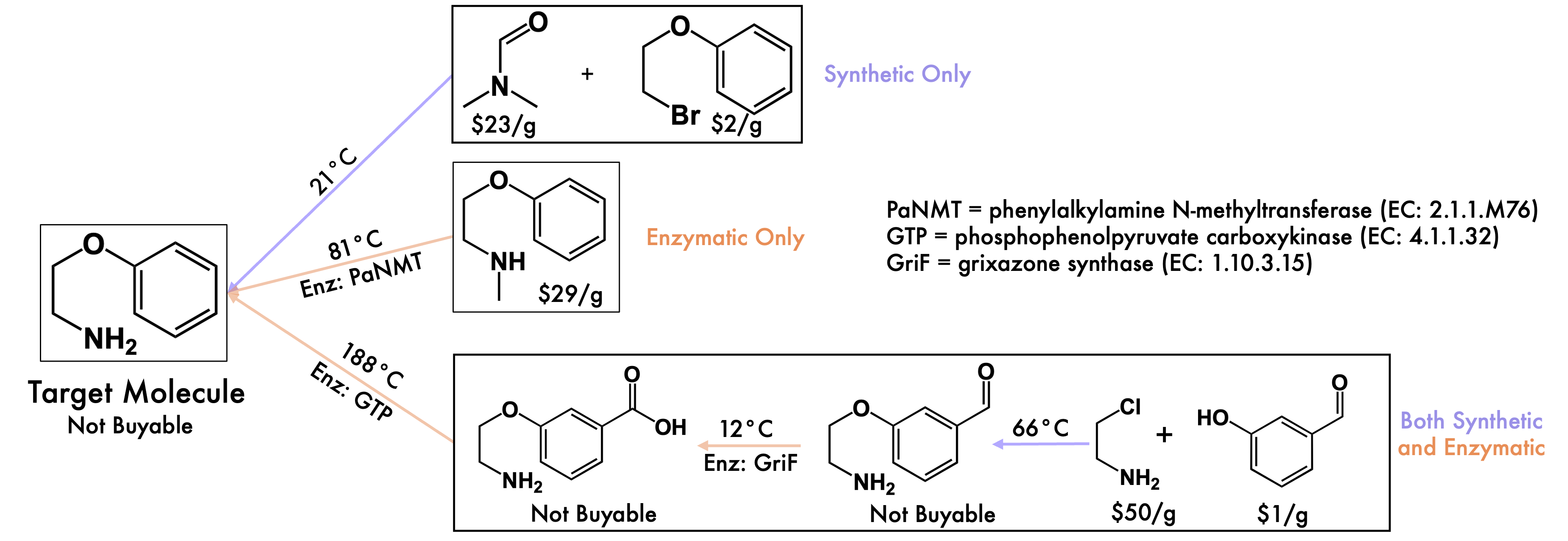}
    \caption{The figure illustrates the enzymatic, synthetic, and hybrid pathways for the synthesis of 2-phenoxyethanamine. The target molecule, marked as ``Not Buyable,'' is shown at the root of the pathway. Pathways are categorized into three types:\\ \textbf{1. Synthetic Only (purple):} Reactions that rely solely on chemical synthesis, involving commercially available starting materials and intermediates.\\ \textbf{2. Enzymatic Only (orange):} Reactions catalyzed exclusively by enzymes such as PaNMT (phenylalkylamine N-methyltransferase) and GTP (phosphoenolpyruvate carboxykinase).\\ \textbf{3. Hybrid Pathways (purple and orange):} Routes that combine enzymatic and synthetic steps to achieve the desired product. \\ Each node represents a compound with its associated cost in USD per gram, while edges denote reaction conditions such as temperature and enzyme usage. This visualization highlights the interplay between synthetic and enzymatic strategies in optimizing cost-effective and feasible pathways for 2-phenoxyethanamine production.}
\end{centering}
\end{figure}

\newpage

\section{Supplementary Figure 5: Hybrid pathway tree for 1-naphthylamine}

\begin{figure}[h!]
\begin{centering}
    \includegraphics[width=\textwidth]{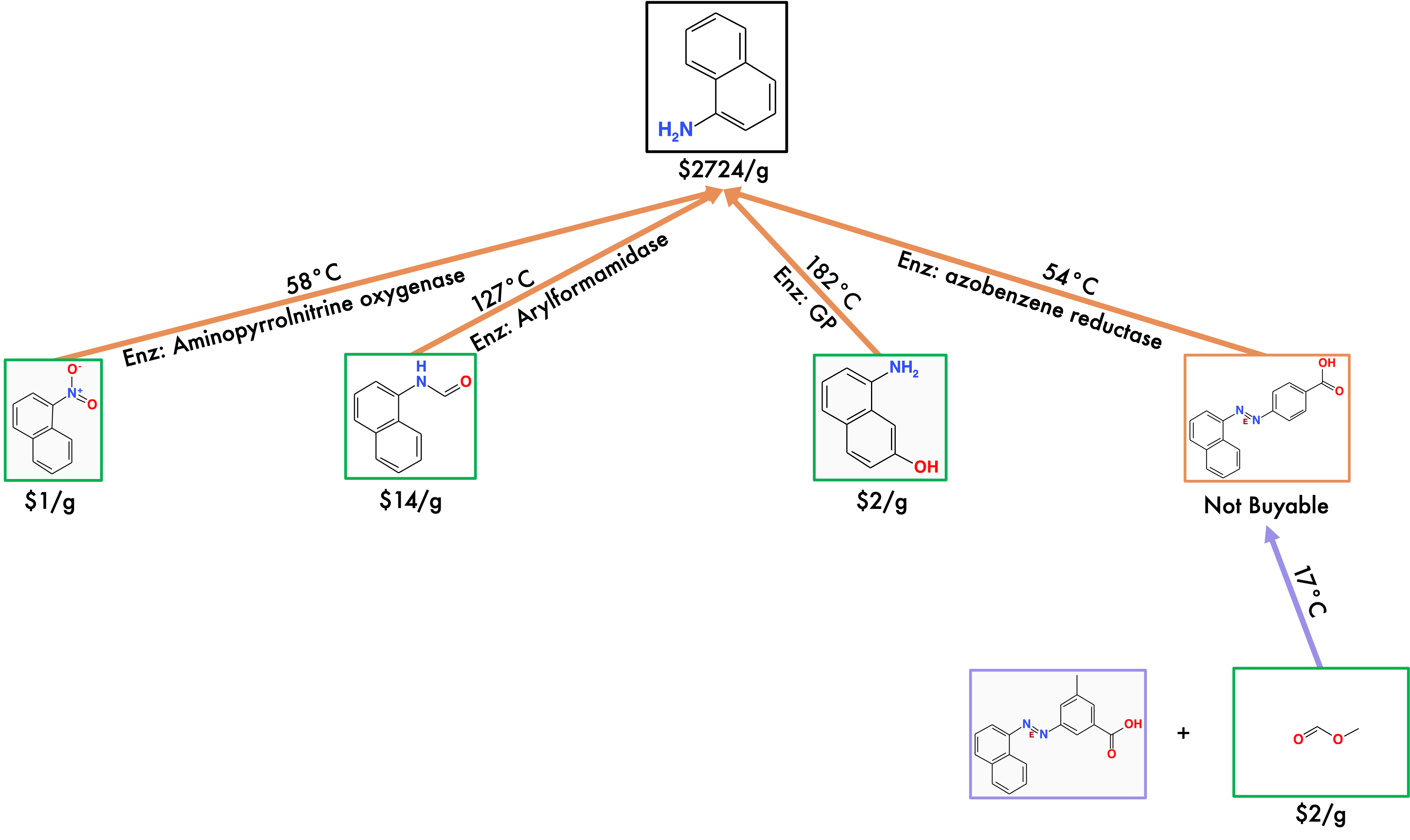}
    \caption{The hybrid pathway tree for 1-naphthylamine highlights potential synthetic and enzymatic routes to produce this compound at a lower cost than its current import price into Canada. The root node represents 1-naphthylamine, with branches depicting enzymatic reaction steps (orange edges) and their associated conditions, including temperature and enzyme type. Intermediate compounds are annotated with their respective costs in USD per gram, with green-bordered nodes indicating commercially available starting materials. \\ This analysis demonstrates feasible pathways that combine enzymatic and synthetic steps to achieve significant cost reductions. For example, enzymatic transformations using aminopyrrolnitrinre oxygenase or azobenzene reductase lead to intermediates priced at \$1/g and \$2/g, respectively. This pathway visualization underscores the economic potential of hybrid approaches to synthesize 1-naphthylamine domestically, reducing reliance on high-cost imports.}
\end{centering}
\end{figure}

\newpage

\section{Supplementary Figure 6: Pathway comparisons with RetroBioCat \cite{finnigan2021retrobiocat}}

\begin{figure}[h!]
\begin{centering}
    \includegraphics[width=\textwidth]{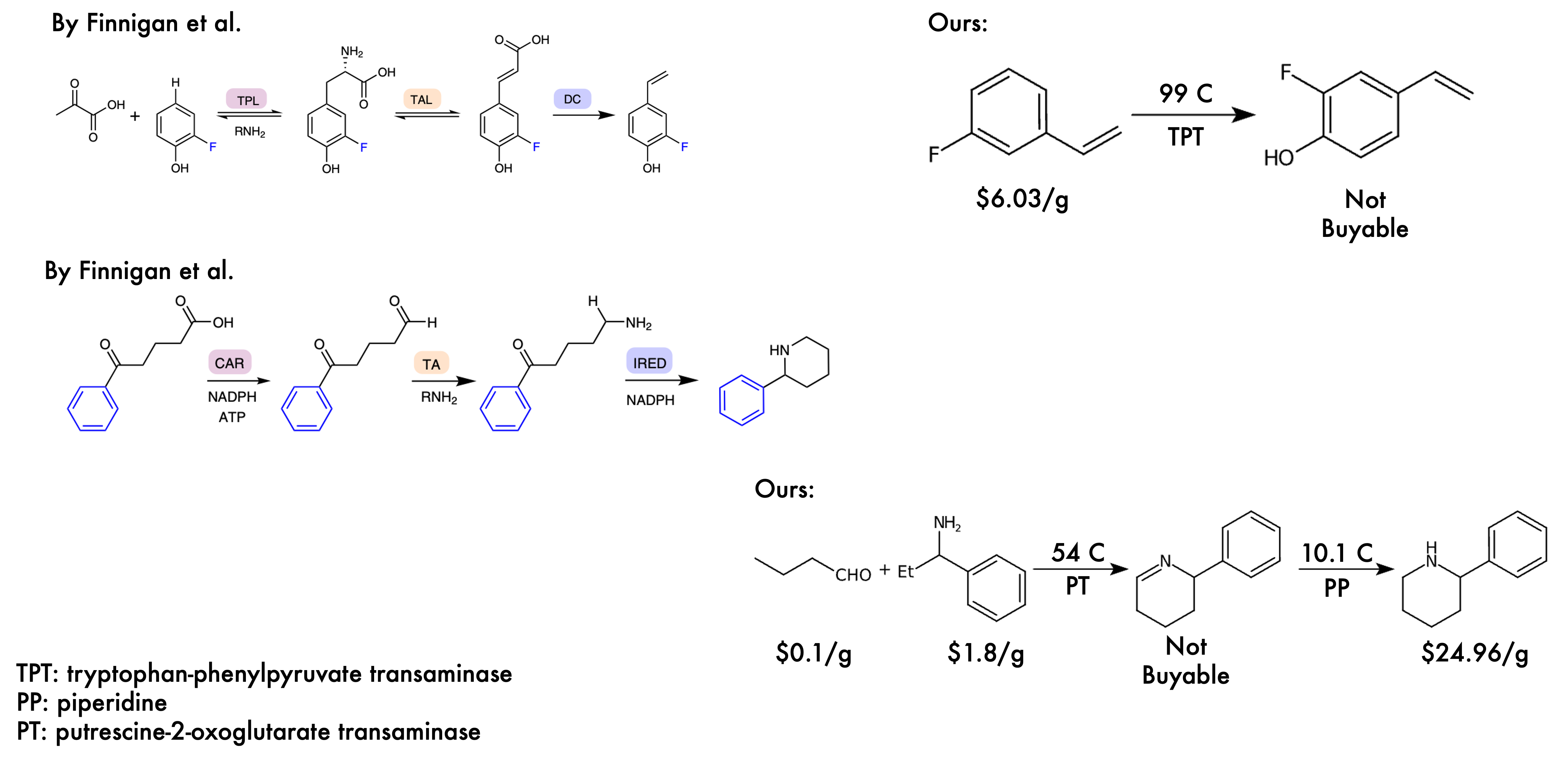}
    \caption{Comparison of biosynthetic pathways generated by RetroBioCat (Finnigan et al.) and our approach for two target compounds.\\ 
    \\ \textbf{Top Panel:} The RetroBioCat pathway for the fluorinated compound involves multiple enzymatic steps, including TPL (tryptophan-phenylpyruvate transaminase), TAL, and DC, leading to the final product. In contrast, our approach achieves the same product using a single enzymatic step with TPT at 99°C, significantly reducing complexity and cost (\$6.03/g).\\
    \textbf{Bottom Panel:} For the amine compound, the RetroBioCat pathway involves three enzymatic steps (CAR, TA, IRED) requiring NADPH and ATP. Our hybrid pathway uses two enzymatic steps: PT (putrescine-2-oxoglutarate transaminase) at 54°C and PP (piperidine transaminase) at 10.1°C. This results in a lower overall cost (\$0.1/g and \$1.8/g).}
\end{centering}
\end{figure}

\newpage

\section{Supplementary Figure 7: Pathway from PaRoutes replicated by MHNpath }

\begin{figure}[h!]
\begin{centering}
    \includegraphics[width=0.9\textwidth]{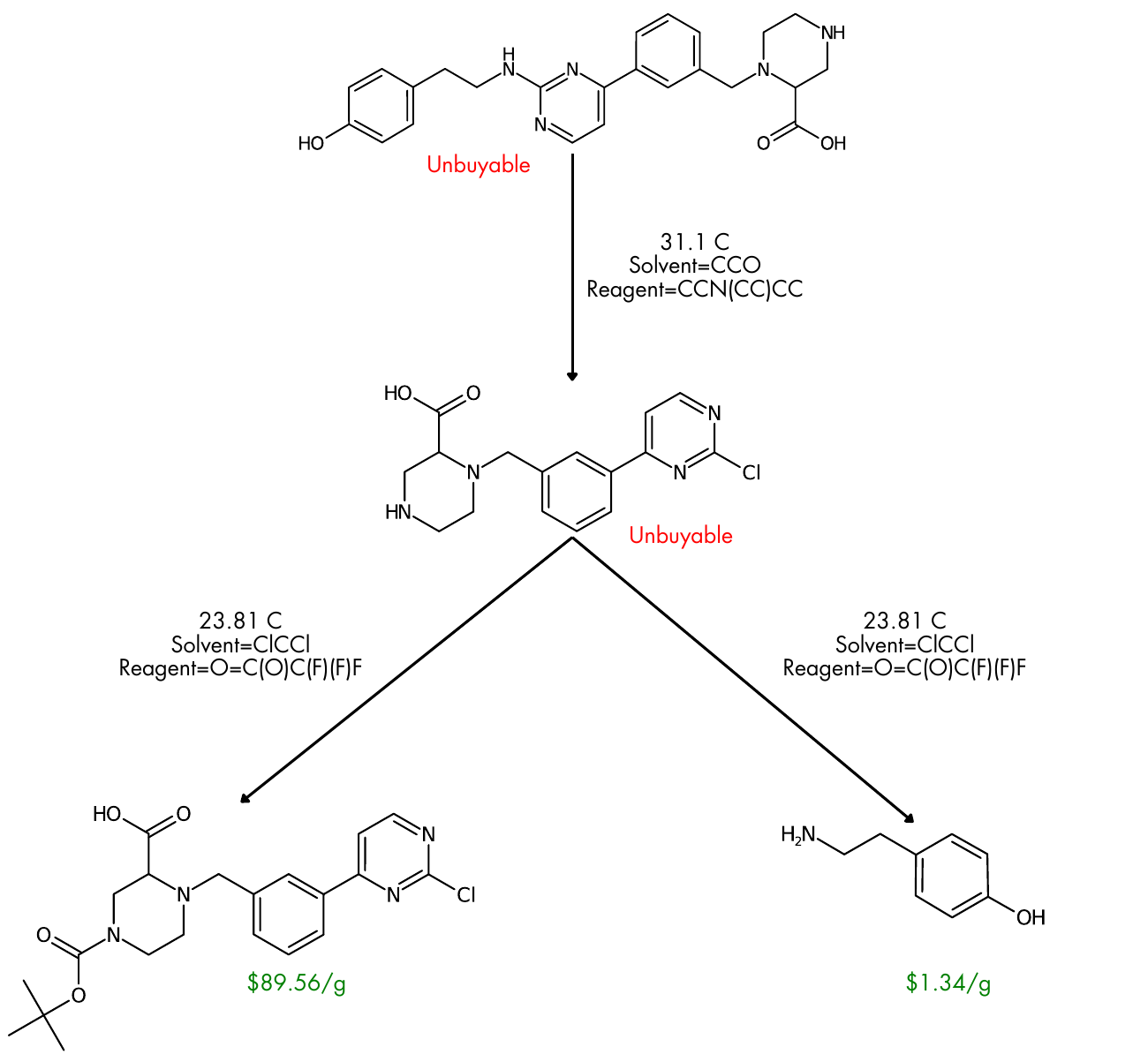}
    \caption{This figure illustrates a synthetic pathway sourced from the PaRoutes dataset, which compiles best-known synthetic routes derived from patents and literature. The pathway shown was successfully replicated by our model, MHNpath. Each node represents a compound, annotated with its associated production cost in USD per gram. Edges represent reaction steps, annotated with reaction conditions such as temperature, solvent, and reagents. \\ 
    The root node corresponds to the target compound which is not buyable. MHNpath accurately reproduces the multi-step pathway, including intermediate compounds and their respective costs (\$89.56/g and \$1.34/g for the final intermediates). This demonstrates the model's ability to replicate established pathways with high fidelity, validating its utility in identifying cost-effective synthetic routes.}
\end{centering}
\end{figure}

\newpage

\section{Supplementary Figure 8: New pathway predicted by MHNpath for arformoterol}

\begin{figure}[h!]
\begin{centering}
    \includegraphics[width=0.8\textwidth]{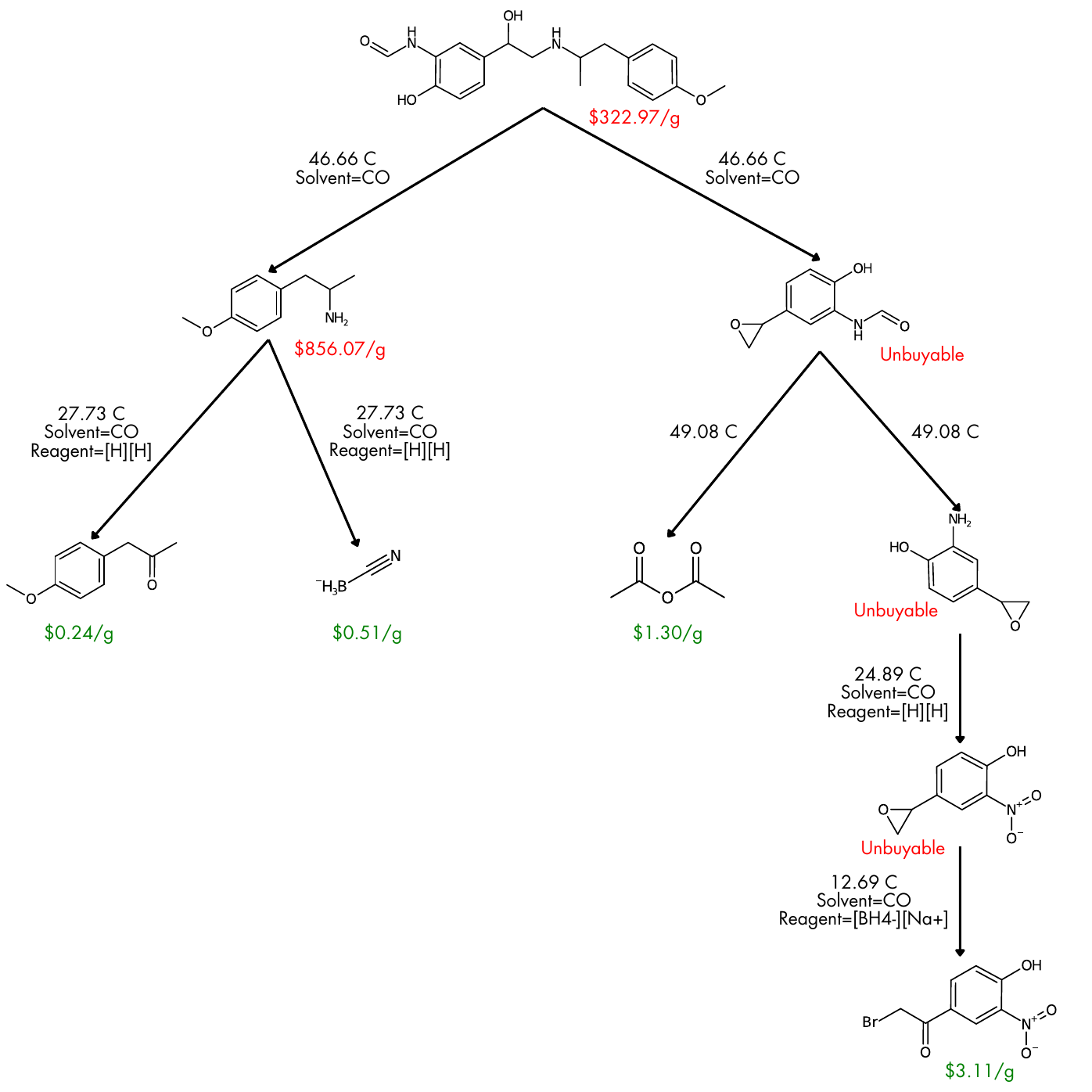}
    \caption{The figure presents a novel pathway for the synthesis of arformoterol predicted by our model, MHNpath. Starting from inexpensive precursors (\$3.11/g, \$1.30/g, \$0.51/g, and \$0.24/g), MHNpath identifies a 4-step pathway that minimizes overall cost while maintaining synthetic feasibility. Each node represents a compound annotated with its associated production cost in USD per gram, and edges denote reaction steps with detailed conditions such as temperature, solvent, and reagents.\\
    In comparison, Levin et al. (\cite{levin2022merging}) proposed a 5-step biocatalytic cascade pathway for arformoterol synthesis involving multiple enzymes (e.g., AMO, LAM, CYP2D6) and cofactors (e.g., NADPH, FAD). While their approach demonstrates the utility of biocatalysis, MHNpath achieves a shorter and more cost-effective route by leveraging hybrid strategies that integrate enzymatic and synthetic steps.}
\end{centering}
\end{figure}

\newpage

\section{Supplementary Figure 9: Alternative pathway predicted by MHNpath for Lupinine \cite{wang2024enantioselective}}

\begin{figure}[h!]
    \centering
    \begin{subfigure}{\textwidth}
        \centering
        \includegraphics[width=0.8\textwidth]{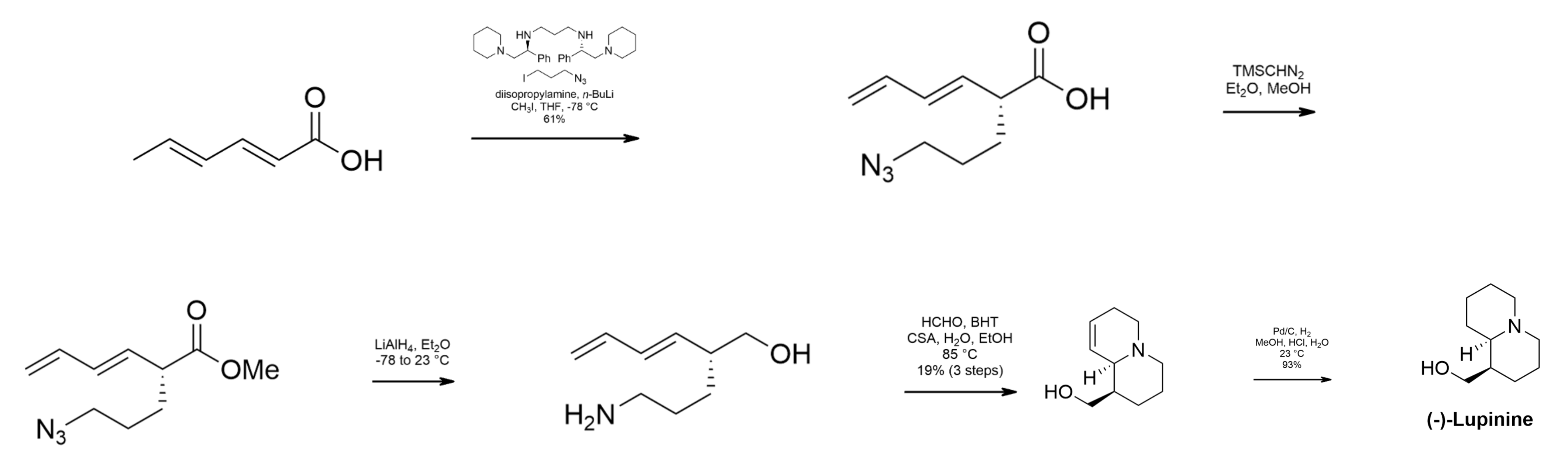}
        \caption{A five-step pathway to synthesize (-)-Lupinine developed by Wang et al. \cite{wang2024enantioselective}, utilizing ChemByDesign's database of newly discovered pathways. This method involves a range of temperatures from $-78\degree$C to $85\degree$C.}
    \end{subfigure}

    \begin{subfigure}{\textwidth}
        \includegraphics[width=\textwidth]{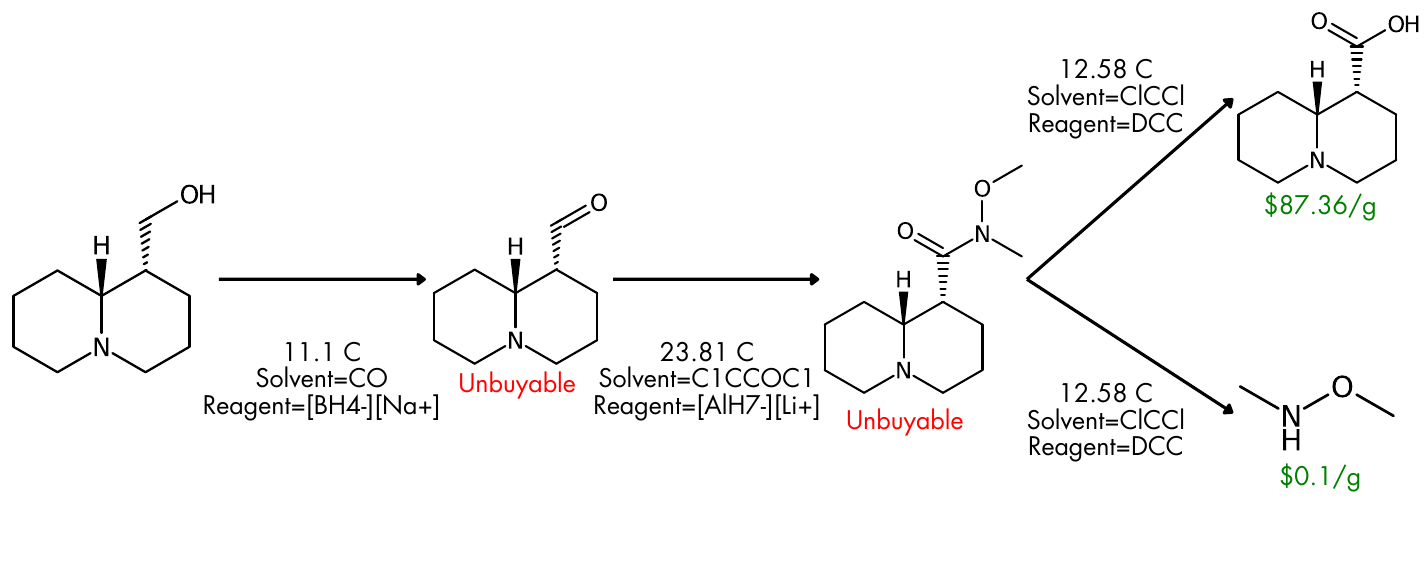}
        \caption{A three-step pathway predicted by the MHNpath model, offering a cost-effective alternative with precursor costs of \$87.36 and \$0.10 per gram. The reaction proceeds under more moderate conditions, with temperatures at $12.58\degree$C, $11.1\degree$C and $-23.81\degree$C, demonstrating improved efficiency and feasibility compared to Wang et al.'s approach.}
    \end{subfigure}
    \caption{Comparison of pathways for Lupinine synthesis}
\end{figure}

\newpage

\section{Supplementary Discussion 1: Output tree structure}

The output tree structure generated during the analysis is designed to represent a hierarchical decomposition of chemical reactions and their associated costs. Each node in the tree represents a chemical compound, while the edges connecting nodes represent chemical reactions that transform one compound into another. This section provides a detailed explanation of the tree structure, how to interpret the text files, and guidance on deprecated fields.

\subsection*{Node Structure}
Each node in the tree is defined as an instance of the \texttt{Node} class with the following attributes:
\begin{itemize}
    \item \texttt{smiles:} A string representing the SMILES (Simplified Molecular Input Line Entry System) notation of the compound.
    \item \texttt{cost\_usd\_per\_g:} The cost of the compound in USD per gram.
    \item \texttt{depth:} The depth of the node in the tree, with the root node having a depth of 0.
    \item \texttt{subtrees:} A list of tuples, where each tuple consists of an \texttt{Edge} object and a child \texttt{Node}. This represents the possible reactions (edges) leading to subsequent compounds (nodes).
\end{itemize}

\subsection*{Edge Structure}
Each edge connecting two nodes is represented as an instance of the \texttt{Edge} class with the following attributes:
\begin{itemize}
    \item \texttt{reaction\_smiles:} A string representing the reaction in SMILES format.
    \item \texttt{temperature:} The temperature (in Kelvin) at which the reaction occurs.
    \item \texttt{enzyme:} An identifier for any enzyme used in the reaction. If no enzyme is used, this value is set to 0.
    \item \texttt{score:} A numerical score associated with the reaction, typically indicating its likelihood or efficiency.
    \item \texttt{rule:} A string representing the transformation rule applied during the reaction.
    \item \texttt{label:} An identifier for labeling purposes.
\end{itemize}

\subsection*{Example Node Representation}
An example node from the text file is shown below:
\begin{verbatim}
{
  "reaction_smiles": "ccc-c1cccc(CN2CCN(C(=O)OC(C)(C)C)CC2C(=O)O)c1.CN(...)",
  "temperature": 300,
  "enzyme": 0,
  "score": -1.00026676,
  "rule": "[#7;a:5]:[c;H0;D3;+0:4](:[#7;a:6])-[NH;D2;+0:8]-[C:7]...",
  "label": 0,
  "type_dis": 0,
  "subtree": {
    "smiles": "ccc-c1cccc(CN2CCN(C(=O)OC(C)(C)C)CC2C(=O)O)c1",
    "cost_usd_per_g": 500,
    "depth": 1,
    "subtrees": []
  }
}
\end{verbatim}

In this example:
\begin{itemize}
    \item The root node represents a compound with SMILES notation \texttt{ccc-c1cccc(CN2CCN(C(=O)OC(C)(C)C)CC2C(=O)O)c1}, a cost of \$500 per gram, and a depth of 1.
    \item The subtree contains no further decomposition (\texttt{subtrees: []}), indicating that it is a leaf node.
\end{itemize}

Certain fields, such as \texttt{type\_dis} or \texttt{buyable}, may appear in some text files. These fields were originally intended for additional functionalities but are now deprecated and should be ignored during analysis.

\subsection*{How to Read and Interpret Text Files}
The text files are structured in JSON format, making them easy to parse programmatically. To reconstruct the tree:
\begin{enumerate}
    \item Parse each JSON object into corresponding \texttt{Node} and \texttt{Edge} instances.
    \item Recursively traverse through each node's \texttt{subtrees} attribute to build the hierarchical structure.
    \item Ignore deprecated fields like \texttt{type\_dis} or \texttt{buyable}.
    \item Sample code to read, write and process these JSON files is given also with our other code.
\end{enumerate}

This tree structure provides an intuitive way to represent reaction pathways and their associated costs, facilitating efficient analysis and decision-making in synthetic chemistry workflows.

\newpage

\section{Supplementary Note 2: Molecules attempted from RetroBioCat \cite{finnigan2021retrobiocat}}

The molecules attempted were
\begin{itemize}
    \item 4-ethenyl-2-fluorophenol
    \item 2-phenylpiperidine
    \item 5-(6-amino-2-fluoro-purin-9-yl)-2-ethynyl-2-(hydroxymethyl)tetrahydrofuran-3-ol
    \item L-Alloisoleucine
    \item 6-hydroxy-4-methylhexanoic acid
\end{itemize}

\end{document}